\documentclass{aastex63}

\usepackage{txfonts}
\usepackage{graphicx}

\accepted {September 10, 2020}
\submitjournal{ApJ}

\shorttitle{The depth and the vertical extent of the energy deposition...}
\shortauthors{Radziszewski et al.}

\begin{document}

\title{The depth and the vertical extent of the energy deposition layer in a medium-class
 solar flare} 

\correspondingauthor{Krzysztof Radziszewski}
\email{radziszewski@astro.uni.wroc.pl}

\author[0000-0003-0310-1598]{Krzysztof Radziszewski}
\affiliation{Astronomical Institute, University of Wroc{\l}aw, 51-622 Wroc{\l}aw, ul. Kopernika 11, Poland}

\author[0000-0003-1853-2809]{Robert Falewicz}
\affiliation{Astronomical Institute, University of Wroc{\l}aw, 51-622 Wroc{\l}aw, ul. Kopernika 11, Poland}

\author[0000-0002-3989-1619]{Pawe{\l} Rudawy}
\affiliation{Astronomical Institute, University of Wroc{\l}aw, 51-622 Wroc{\l}aw, ul. Kopernika 11, Poland}

\begin{abstract}

We analyze here variations of the position and the vertical extent of the energy deposition layer (EDL) in the C1.6 \textit{GOES}-class solar flare observed at 10:20 UT on 2012 September 10. The variations of the EDL are contrasted with the variations of the spectra and emission intensities recorded in the H$\alpha$ line with the very high time resolution using the MSDP spectrograph at Bia{\l}k{\'o}w Observatory.

The flare radiated hard X-rays (HXR) detectable up to a energy of 70~keV. A numerical model of the flare used in the analysis assumes that the non-thermal electrons (NTEs) carried the external energy to the flare. The NTEs energy flux was derived from a non-thermal component seen in \textit{RHESSI} spectra. The main geometrical parameters of the flare were derived using restored \textit{RHESSI} imaging data.

We found that the variations of the X-ray fluxes recorded in various energy bands and the variations of the H$\alpha$ intensities were well correlated in time during the pre-impulsive and impulsive phases of the flare and they agreed with the variations of the calculated position and vertical extent of the EDL. The variations of the emission noticed in various parts of the H$\alpha$ line profile were caused by individual episodes of energy deposition by the beams of NTEs of various energy spectra on various depths in the chromospheric plasma. These results supplement our previous findings for the solar flare on 21 June 2013, having nearly the same \textit{GOES}-class of C1.1 but HXR emission below 34~keV only \citep{2017ApJ...847...84F} (hereafter Paper I).

\end{abstract}


\section{Introduction} \label{sec:introd}
Solar flares are violent and powerful phenomena that alternate properties of the whole surrounding solar atmosphere, influence properties of the interplanetary environment via various direct and indirect mechanisms, and shape so-called \textit{space weather} and numerous geophysical processes. For this reason, solar flares have been thoroughly observed and analyzed for many dozens of years. Very intensive and effective investigations started in the 1960s when space-based X-ray and EUV data became available.

Solar flares encompass a very complicated system of interrelated processes, taking place in a violently evolving environment of non-potential magnetic fields in the active regions \citep{2013shin.confE..21C, 2011SoPh..271...57J, 2019LRSP...16....3T, 2019ApJ...878..135K}. Solar flare radiation range from X-rays to the radio band, covering almost the entire range of the electromagnetic spectrum \citep{2011SSRv..159...19F}. In the case of the most powerful flares, gamma-emission is also recorded \citep{2005ExA....20...65S}. The time and spatial scales of the physical processes in solar flares are very small, still below the capabilities of present space- and ground-based instruments. New insight into solar flares will be provided by NASA Parker Solar Probe and ESA-NASA Solar Orbiter \citep{2013SoPh..285...25M, 2016SSRv..204....7F}, and ground-based 4-m class telescopes Daniel K. Inouye Solar Telescope (DKIST) in Hawaii, and European Solar Telescope (EST) due to be installed in the Canary Islands \citep{2014SPIE.9147E..07E, 2019AdSpR..63.1389J}. In the meantime, very small-scale processes can only be investigated indirectly, with state-of-the-art numerical models and refined theoretical studies.

The general 2D model of the physical mechanisms and energy release processes occurring in solar flares (so-called CSHKP model), was elaborated more than 50 years ago by \citet{1964NASSP..50..451C}, \citet{1966Natur.211..695S}, \citet{1974SoPh...34..323H}, and \citet{1976SoPh...50...85K}. After numerous modifications over the years, it evolved into a commonly accepted standard model of the flares \citep{1989SoPh..121...75D, 1999spro.proc..381S}. The model assumes that energy accumulated already in non-potential magnetic fields is released via avalanches of local reconnections and then is converted into the kinetic energy of the charged particles, in particular non-thermal electrons, internal, kinetic and potential energy of the plasma, and also cause various electromagnetic emissions. The beams of the non-thermal electrons are guided by loop-like magnetic structures rooted toward the chromosphere and even upper photosphere, where the energy is deposited, causing impulsive local heating and evaporation of chromospheric material.

Although the time and spatial scales cannot be observed, the general characteristics of the processes can be derived from their emissions recorded in hard X-ray (HXR via bremsstrahlung), soft X-ray (SXR; thermal), UV and even visible domains. In the case of visible-domain emissions, the chromospheric emission in the hydrogen H$\alpha$ line ($\lambda$~=~6563~{\AA}) is used as an effective diagnostic tool while its emission depends on the energy deposition in the chromosphere \citep{1973ApJ...184..605V, 1981ApJS...45..635V} and reveals the local configuration of the magnetic fields. Also other spectral lines of the visible range - like the H$\beta$ line ($\lambda$~=~4861~{\AA}) - can be used to study the energy deposition in the lower solar atmosphere \citep{2020ApJ...896..120K}.

High time-resolution spectral observations of the H$\alpha$ line profile provide very valuable data \citep{1988ApJ...324..582Z, 1992LNP...399..372G, 1994SoPh..152..393H, 2017NatCo...815905D, 2007A&A...461..303R, 2017ApJ...847...84F}. \citet{1984ApJ...282..296C} investigated the effect of a non-thermal electron beam and coronal pressure enhancement on synthesized H$\alpha$ line profiles and suggested that increased pressure can cause line profile broadening and increase the total intensity of the H$\alpha$ line. Theoretical analyses were also performed by \citet{1991SoPh..135...65H}. He estimated the cooling time of the chromosphere to be of the order of one second. Thus, structures seen in strong chromospheric lines should change their brightness within similarly short time intervals, clearly revealing a high correlation of the variable HXR and H$\alpha$ emissions. Many observations in the H$\alpha$ line and X-rays have been made. Starting in the early 1980s, Kurokawa and his co-workers observed solar flares through a narrow-band H$\alpha$ filter \citep{1986Ap&SS.118..149K, 1988PASJ...40..357K, 1990SoPh..125..321K}. They found that the time lags between H$\alpha$ and either hard X-ray or microwave emission are shorter than 10 seconds. \citet{2000ApJ...542.1080W} made H$\alpha$ observations of solar flares with a time resolution of 0.033 s. They found that during the seven-second-long period the H$\alpha$ emission of a flare kernel showed fast (0.3--0.7 s) fluctuations correlated with variations of the HXR flux. \citet{2000A&A...356.1067T} found a strong time correlation between HXR and H$\alpha$ emission from a \textit{GOES} X1.3 class flare on 1991 March 13. Radziszewski with co-workers \citep{2006AdSpR..37.1317R, 2007A&A...461..303R, 2011A&A...535A.123R} subsequently investigated time variations of the H$\alpha$ and X-ray emissions observed during the pre-impulsive and impulsive phase of numerous solar flares and showed that in many cases the variations of the X-ray light curves recorded in various energy bands and the variations of the H$\alpha$ light curves and line profiles were highly correlated and the time delay between relevant individual local peaks of the HXR and H$\alpha$ emissions is of order 1-3 s only.

Several investigations have also shown the blue asymmetry of the H$\alpha$ line profile during the early rise phase of the flare (see \citet{1994SoPh..152..393H} and references therein). However, the underlying physical processes remain elusive. In particular, although \citet{1994SoPh..152..393H} proposed that the early blue-shift in the H$\alpha$ line to be due to electron beams with return current, \citet{2015ApJ...813..125K} emphasized the role of steep gradient in the velocity of emitting plasma in shifting the wavelength of maximum opacity, thus adding another dimension of complexity in the implications obtained merely by the shift in the H$\alpha$ line profile. A direct comparison of the observed and modeled properties of the time variations of the H$\alpha$ and X-ray emissions can also give new clues on physical processes acting at the flaring loop footpoints. Falewicz and co-workers \citep{2017ApJ...847...84F} compared time variations of the H$\alpha$ and X-ray emissions observed during the pre-impulsive and impulsive phases of the C-class solar flare with those of plasma parameters and synthesized X-ray emission from a hydrodynamic numerical model of the flare. The time variations of the X-ray fluxes in various energy bands, H$\alpha$ intensities, and H$\alpha$ line profiles were well correlated, and the timescales of the observed variations agreed with the calculated variations of the plasma parameters at the flaring loop footpoint, reflecting the time variations of the vertical extent of the energy deposition layer.

The high-time resolution observations of H$\alpha$ line profiles and light-curves provide comprehensive information on the dynamical processes during the various stages of the flare evolution, which can be compared with numerical models of the flares based on multi-band (X-rays, UV, visual) data and state-of-the-art numerical codes. In this paper, we investigate and discuss the evolution of a so-called energy deposition layer (EDL) of a medium-class C1.6 solar flare observed at 10:20 UT, 2012 September 10, in active region NOAA~11564. In section 2 we present the basic properties of this flare and the data used in this study. In section 3 we describe an analysis of the data. In section 4 we present an extended analysis of fast reactions of the H$\alpha$ emission and shape of the line profile on the impulsive heating of the plasma by the NTEs. Section 5 presents the numerical model of the flare. Section 6 provides the outcomes of a single-loop one-dimensional hydrodynamic (1D-DH) model of the flare in the context of the evolution of the EDL during the various phases of the flare, compared with a detailed analysis of variations of the HXR and H$\alpha$ emissions. In Section 7 we present a discussion and conclusions from this analysis.

\section{C1.6 Solar Flare on 2012 September 10} \label{sec:flare}

\begin{figure}[ht!]
\figurenum{1}
\begin{center}
\includegraphics[width=18.0cm]{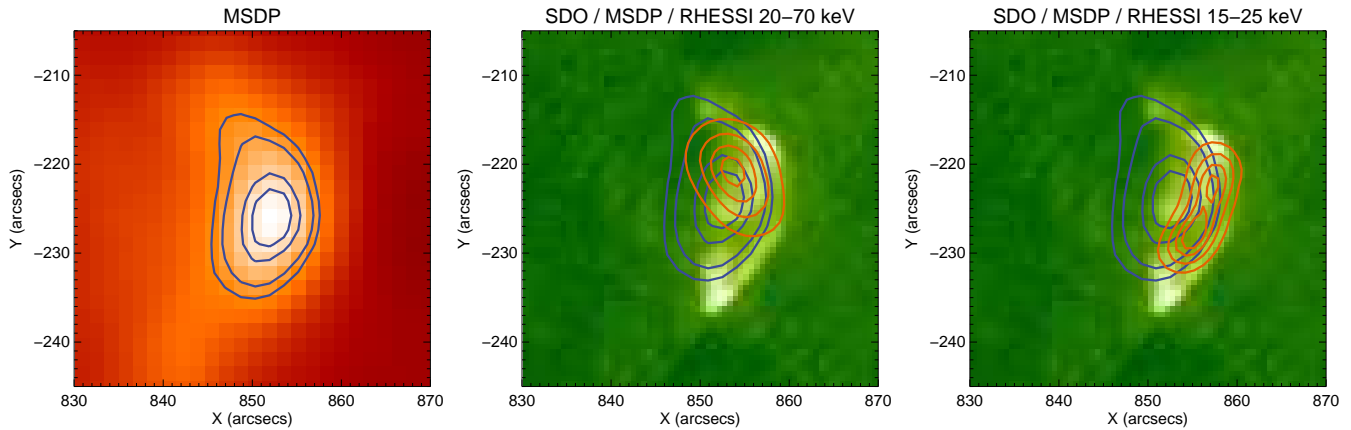}
\end{center}
\vspace{-0.3 cm}
\caption{The C1.6 \textit{GOES}-class solar flare on 2012 September 10, in the active region NOAA~11564. Images: the H$\alpha$ line center image recorded with the \textit{HT-MSDP} system at 10:21:52 UT (left panel) and the 94~{\AA} image recorded by the AIA/\textit{SDO} at 10:21:49 UT (central and right panel). Isocontours: the emission in the H$\alpha$ line center is drawn at 75$\%$ , 80$\%$ , 90$\%$  and 95$\%$  of the maximum of the signal (blue), the hard X-ray emission in the 20--70~keV energy range recorded by the \textit{RHESSI} satellite between 10:21:26 UT and 10:21:54 UT drawn at 30$\%$, 50$\%$, 70$\%$ and 90$\%$ of the peak of the flux (orange in central panel), and the hard X-ray emission in the 15--25~keV energy range between 10:22:16 UT and 10:22:36 UT drawn at 30$\%$, 50$\%$, 75$\%$  and 85$\%$  of the peak of the flux (orange in right panel).
\label{Fig01}}
\end{figure}

\begin{figure}[ht!]
\figurenum{2}
\begin{center}
\includegraphics[width=10.0cm]{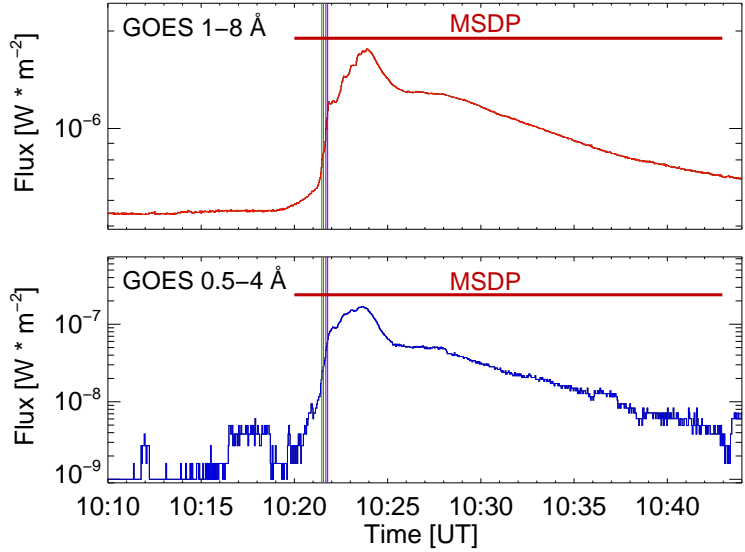}
\end{center}
\vspace{-0.3 cm}
\caption{\textit{GOES-15} light curves recorded in 1--8~{\AA} band (upper panel) and 0.5--4~{\AA} band (lower panel) during the C1.6 \textit{GOES}-class solar flare on 2012 September 10. Red horizontal lines show the time interval when observations were collected with the \textit{MSDP} spectrograph in Bia{\l}k{\'o}w. Double green and purple vertical lines indicate four intermediate maxima of the hard X-ray emission recorded by \textit{RHESSI} satellite above 10~keV (see main text for details).
\label{Fig02}}
\end{figure}

\begin{figure}[ht!]
\figurenum{3}
\begin{center}
\includegraphics[width=9.0cm]{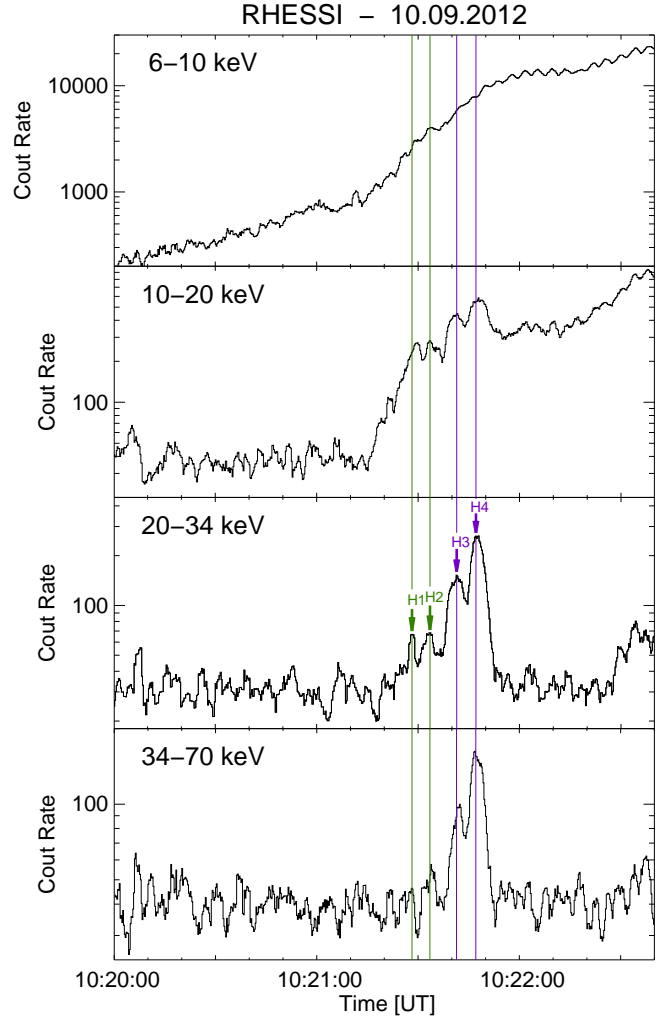}
\end{center}
\vspace{-0.3 cm}
\caption{X-ray light curves recorded by \textit{RHESSI} satellite in four energy ranges: 6--10~keV, 10--20~keV, 20--34~keV, and 34--70~keV during the impulsive phase of C1.6 \textit{GOES}-class solar flare on 2012 September 10. The light curves were recorded without the attenuators (A1 was activated at 10:22:44 UT), demodulated to a time resolution of 250 ms and smoothed with a one-second-wide boxcar filter. Vertical green lines (marked H1, H2) and purple lines (marked H3, H4) indicate four intermediate maxima of the hard X-ray emission observed in the 20--34~keV energy range (see main text for discussion).
\label{Fig03}}
\end{figure}

The C1.6 \textit{GOES}-class solar flare occurred in the $\beta$$\gamma$-type active region NOAA~11564 (S13W58, x~=~850$"$, y~=~-225$"$) at about 10:20 UT on 2012 September 10 (Fig.~\ref{Fig01}). A very faint pre-flare increase of the soft X-ray emission was recorded by \textit{GOES-15} satellite at about 10:19 UT. The initial increase of the X-ray emission started at 10:19:30 UT, the impulsive phase of the flare started at 10:21:15 UT and ended at 10:21:52 UT (Fig.~\ref{Fig02}). Maximum emission in GOES was C1.6 at 10:24:00 UT (Fig.~\ref{Fig02}). Between 10:21:25 UT and 10:21:52 UT, the \textit{RHESSI} satellite \citep{2002SoPh..210....3L} recorded a strong hard X-ray emission in the 20--70~keV energy range (Fig.~\ref{Fig03}). The X-ray flux was low enough that the attenuators were activated only after 10:22:40 UT (before this time they remained in A0 state).


\begin{figure}[ht!]
\figurenum{4}
\includegraphics[width=5.4cm]{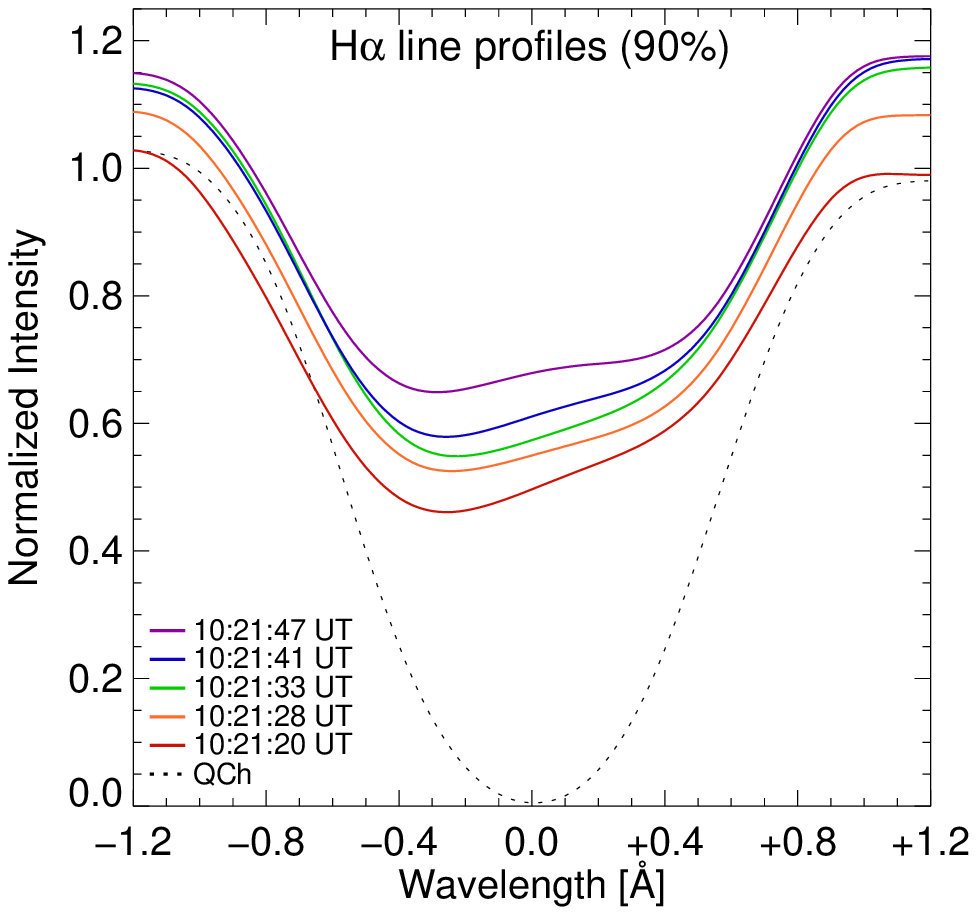}
\includegraphics[width=6.2cm]{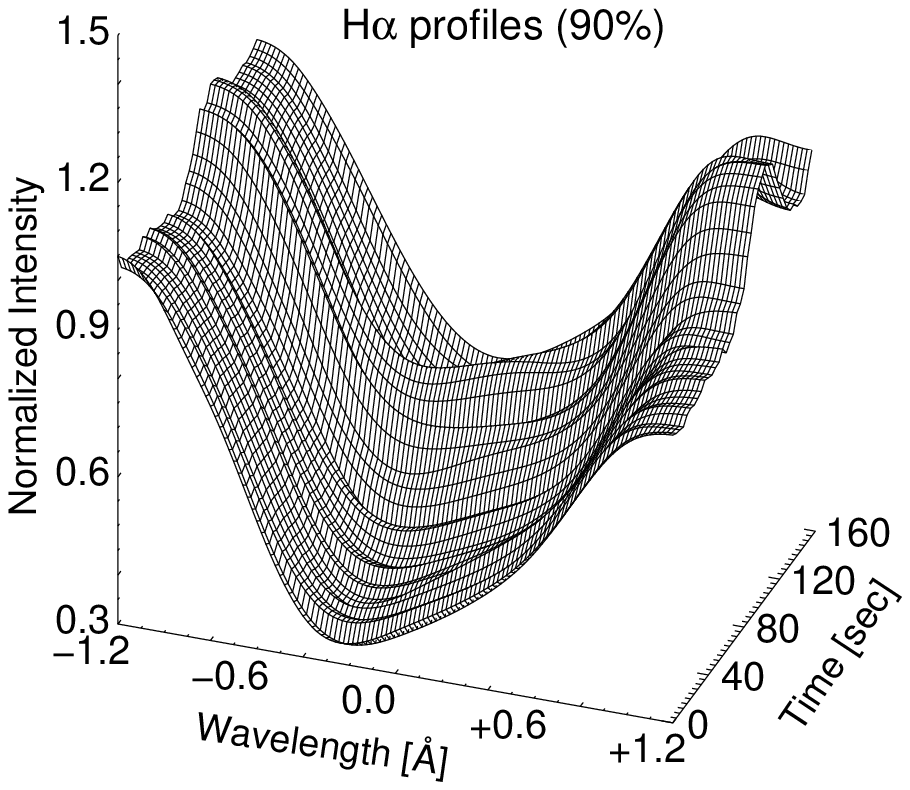}
\includegraphics[width=5.4cm]{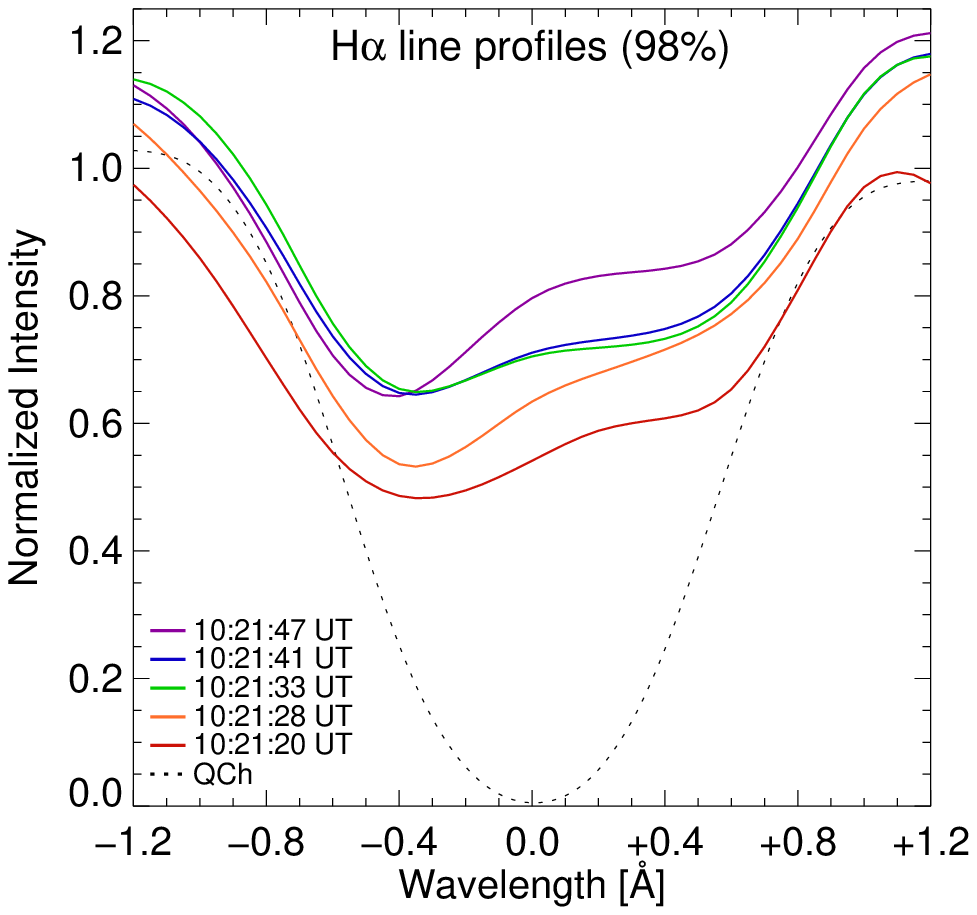}
\vspace{-0.3 cm}
\caption{Left panel: variations of the mean H$\alpha$ line profile of the flaring kernel delimited with an isocontour of 90$\%$ of
the highest intensity in the H$\alpha$ line center. The mean H$\alpha$ line profile of the nearby region of the quiet solar chromosphere (QCh) is shown as the dotted black line. Profiles drawn in orange, green, blue, and purple correspond to the H1, H2, H3 and H4 peaks of the HXR recorded at 10:21:28 UT, 10:21:33 UT, 10:21:41 UT, and 10:21:47 UT, respectively. The red profile was taken during the impulsive phase of the flare at 10:21:20 UT (before the H1 peak). Middle panel: full time cadence series of the mean H$\alpha$ line profiles of a part of the flaring kernel delimited with the isocontour of 90$\%$ (as in the left panel). Time is given in seconds after 10:21:20 UT. Right panel: variations of the mean H$\alpha$ line profile of the brightest fragment of the flaring kernel delimited with an isocontour of 98$\%$ of the largest intensity in the H$\alpha$ line center. All profiles are normalized to unity from line center intensity of QCh to blue wing at -1.0~{\AA}.
\label{Fig04}}
\end{figure}

Between 10:20 UT and 10:43 UT, the flare was also observed in the hydrogen H$\alpha$ line ($\lambda$~=~6563~{\AA}) with the Horizontal Telescope (HT) and Multi-channel Subtractive Double Pass imaging spectrograph (MSDP) installed at the Bia{\l}k{\'o}w Observatory of the University of Wroc{\l}aw \citep{1991A&A...248..669M, 1993BR}. The field of view (FOV) of the HT-MSDP compound system covered an area of 942 x 119 arcsec\textsuperscript{2} on the Sun, centered on H$\alpha$ brightenings seen in the NOAA~11564. Two-dimensional spectra-images formed by the nine-channel prism box covered a 2.4~{\AA} wide wavelength band around the H$\alpha$ line and were recorded with the fast Andor iXon3 885 camera (1002 x 1004 px\textsuperscript{2}) with a spatial resolution of 1.6 seconds of arc per hardware pixel and spectral resolution of 0.4~{\AA}. The cadence of the exposures was equal to 0.05 seconds (equal to 20 spectra-images per second). The time variations of the H$\alpha$ line profiles averaged over the brightest part of the H$\alpha$ structure delimited with an isocontour of 90$\%$  of the highest signal are shown in Figure~\ref{Fig04}.

\begin{figure}[ht!]
\figurenum{5}
\begin{center}
\includegraphics[width=9.0cm]{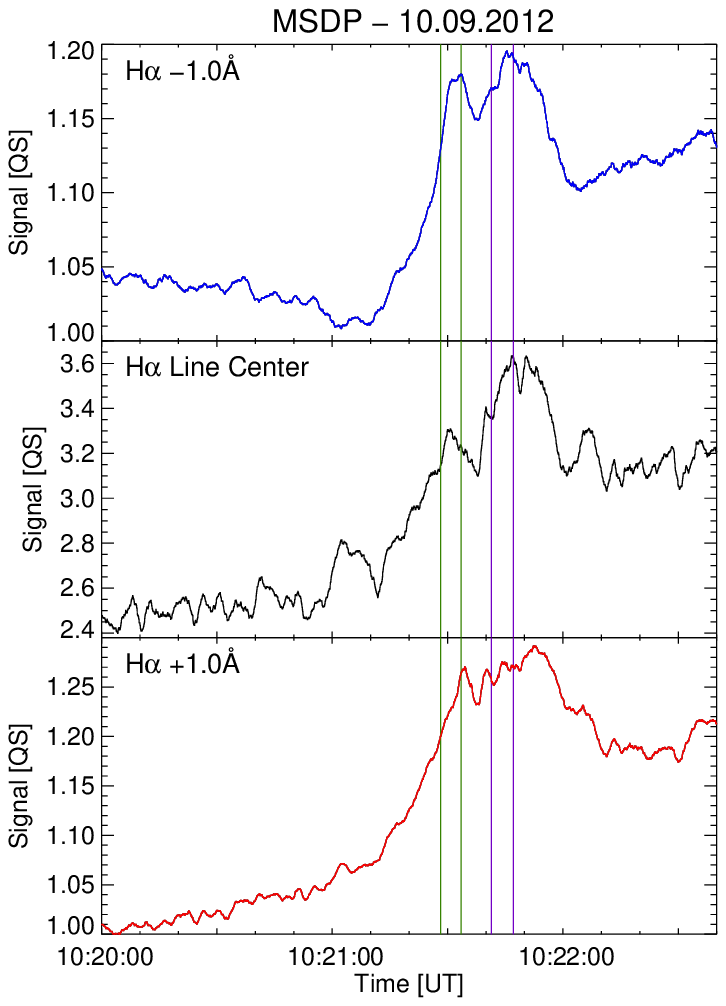}
\end{center}
\vspace{-0.3 cm}
\caption{H$\alpha$ line light curves recorded in $\lambda$~=~H$\alpha_0$-1.0~{\AA}, $\lambda$~=~H$\alpha_0$ and $\lambda$~=~H$\alpha_0$+1.0~{\AA} with the \textit{HT-MSDP }system during the C1.6 \textit{GOES}-class solar flare on 2012 September 10. Double green and purple vertical lines indicate four intermediate maxima (H1--H4) of the hard X-ray emission recorded by the \textit{RHESSI} satellite above 20~keV.
\label{Fig05}}
\end{figure}

The H$\alpha$ emission of the flare was correlated in space and in time with the impulsive brightenings recorded in the hard X-ray and UV bands by the \textit{RHESSI} and the \textit{SDO} satellites \citep{2002SoPh..210....3L, 2012SoPh..275....3P}, and it reached a maximum during the impulsive phase of the flare. The four intermediate emission peaks H1-H4 in the hard X-rays at 10:21:28 UT, 10:21:33 UT, 10:21:41 UT, and 10:21:47 UT, respectively, coincided with the intermediate maxima of the emission in the H$\alpha$ line (Fig.~\ref{Fig05}). Two intermediate X-ray peaks, marked H1 and H2, marked with two green vertical lines, coincide with the pronounced H$\alpha$ peak observed at about 10:21:30 UT. The next two X-ray peaks, marked H3 and H4, coincided with the second, longer-lasting and sub-structured peak of the H$\alpha$ emission observed between 10:21:35 UT and 10:22:05 UT. All intermediate increases of the H$\alpha$ emission were recorded in various parts of the H$\alpha$ line profile occurred immediately after or nearly simultaneously with the hard X-ray pulses. Some differences in variations recorded in various parts of the H$\alpha$ line profile reflect differences in penetration depths of the individual NTEs beams. The time derivatives of the \textit{GOES} light curves show three unambiguous local maxima coinciding with the H1, H3 and H4 peaks seen in the \textit{RHESSI} data. The H2 peak seen in the \textit{RHESSI} data does not have a clear counterpart in the derivatives of the \textit{GOES} light curves, but a small increase of the derivative of the 0.5--4~{\AA} \textit{GOES} light curve can be discerned. A long and dense H$\alpha$ surge was ejected along an arcade of the magnetic loops anchored in the NOAA~11564 active region just at the beginning of the impulsive phase of the flare. However, its optically dense plasma did not obscure any H$\alpha$ flaring kernel up to 10:22:40 UT.


\section{Data processing} \label{sec:data}

\begin{figure}[ht!]
\figurenum{6}
\begin{center}
\includegraphics[width=5.6cm]{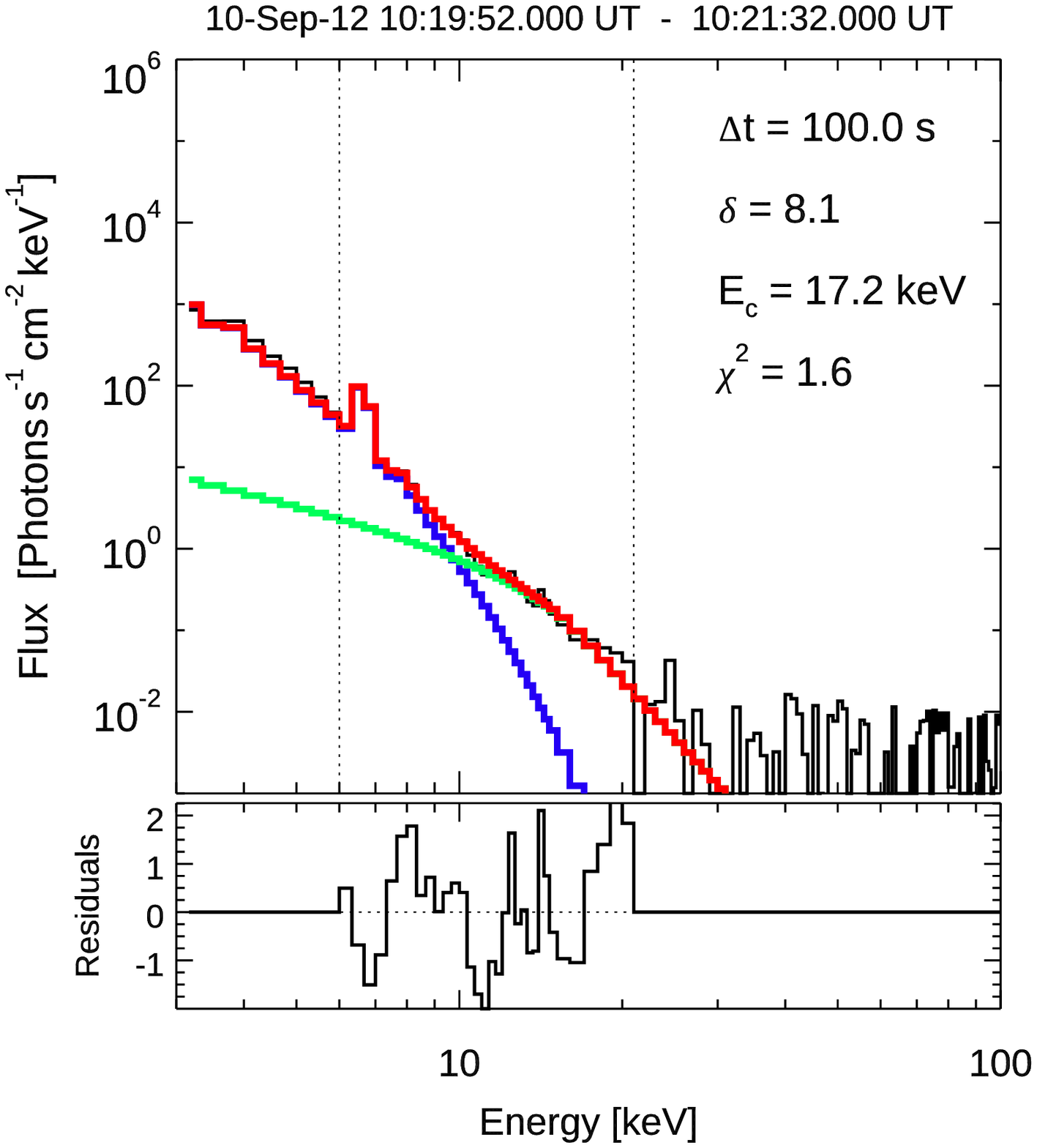}
\includegraphics[width=5.6cm]{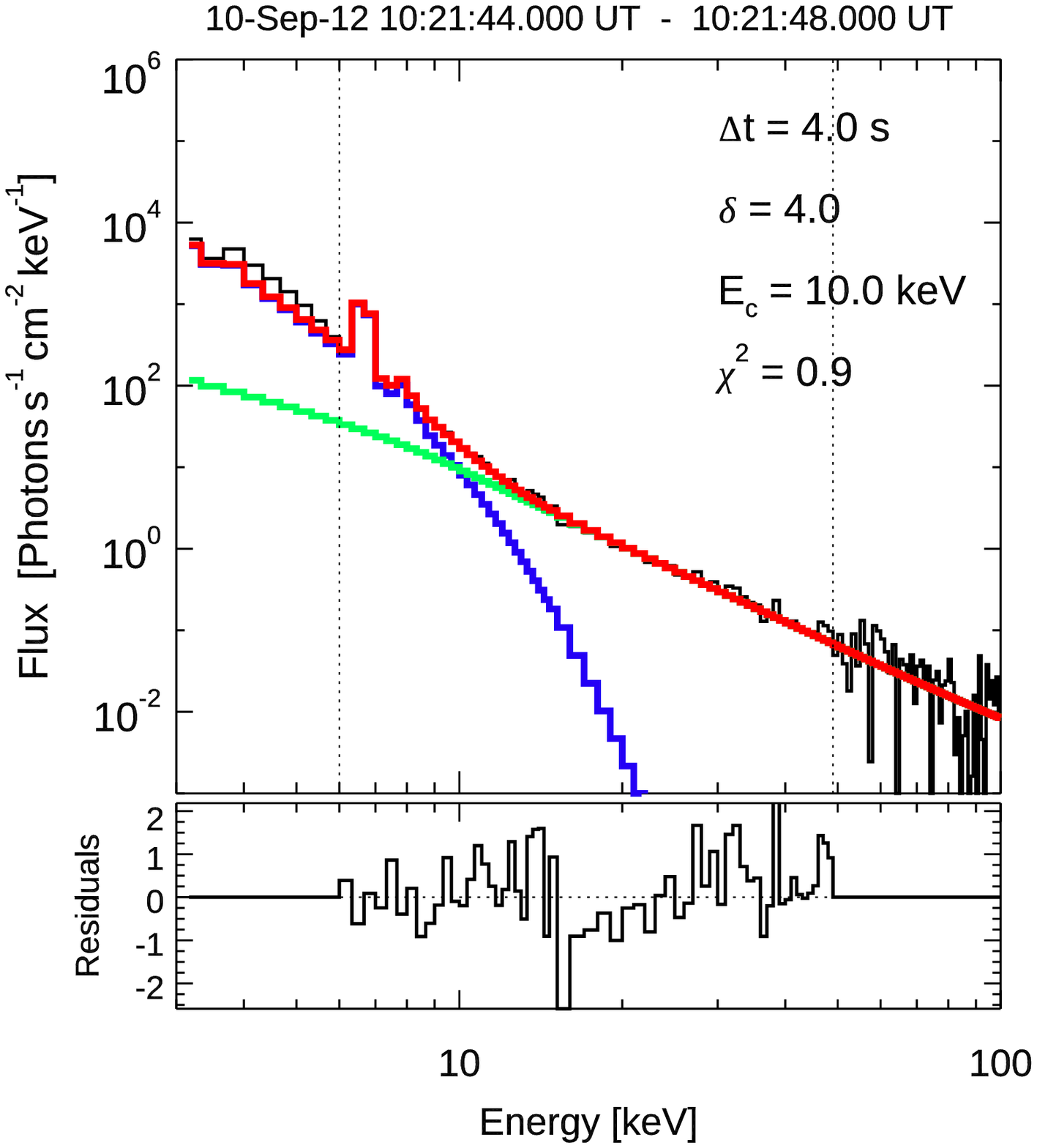}
\includegraphics[width=5.6cm]{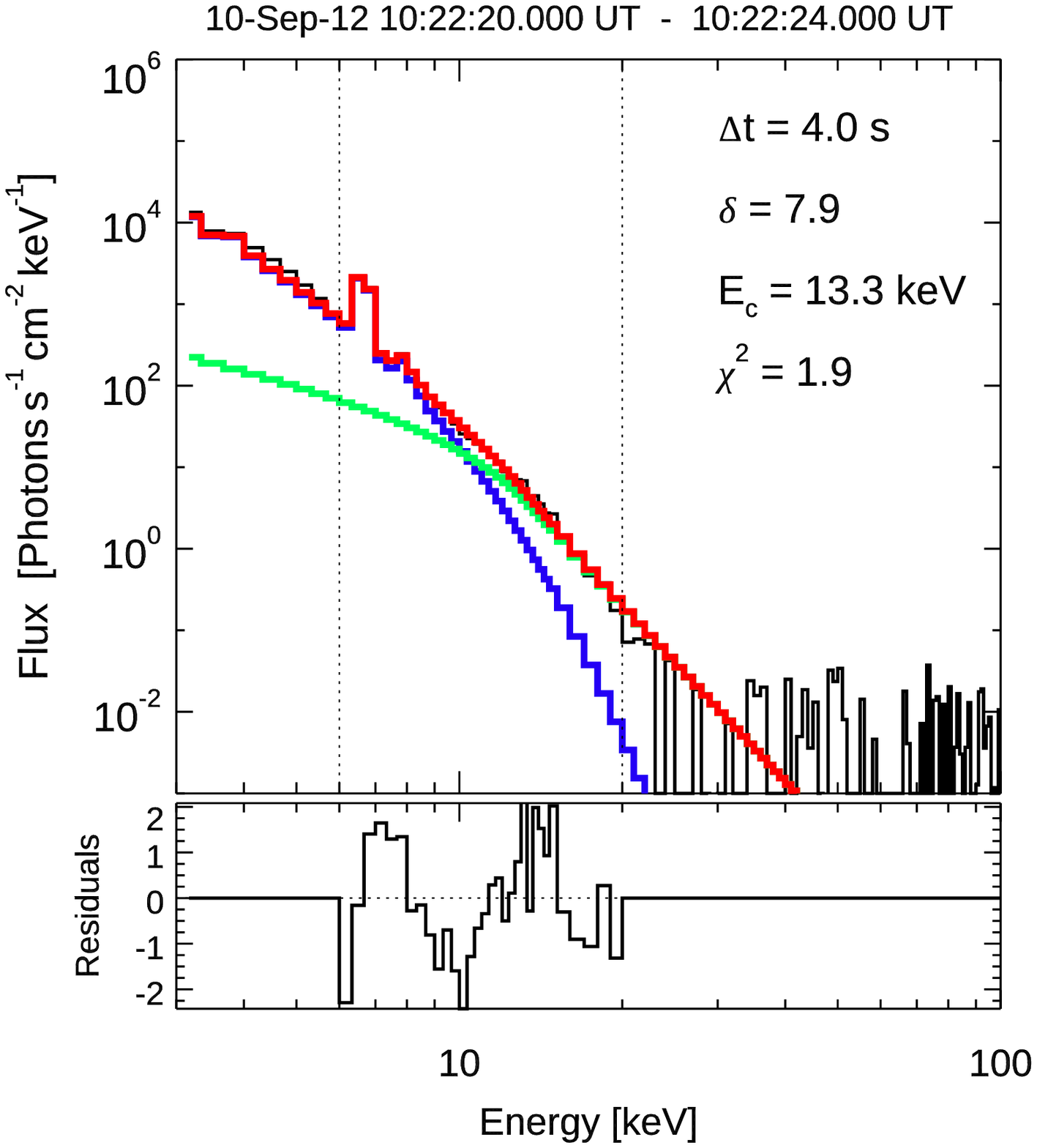}
\end{center}
\vspace{-0.3 cm}
\caption{X-ray emission recorded by the \textit{RHESSI} satellite during the pre-impulsive phase (upper left panel), the impulsive phase (upper central panel), and at the end of the impulsive phase (upper right panel) of the C1.6 \textit{GOES}-class solar flare on 2012 September 10. Blue and green curves show the fits calculated using a single-temperature thermal model and a thick-target model, respectively. Red lines show the sums of both fits. Dotted vertical lines delimit energy ranges used for fitting. Residuals of the fits are shown below the spectra. E\textsubscript{c} and $\delta$ parameters were calculated using the thick-target model and the parametrized models of the injected NTE beams.
\label{Fig06}}
\end{figure}

Data collected by the \textit{RHESSI} with its 3F, 5F, 6F, and 8F detectors allowed the restoration of images of the flaring structures seen in X-rays. The PIXON image restoration algorithm was applied, with effective pixel size equal to one second of arc. A 20-second-long signal accumulation time was selected to obtain a sufficient signal-to-noise ratio \citep{1996ApJ...466..585M, 2002SoPh..210...61H}. X-rays data from \textit{RHESSI} were applied also for the reconstruction of light curves and X-ray energy spectra (Fig.~\ref{Fig06}). X-ray fluxes recorded in the 6--10~keV energy range were summed over the front segments of six detectors: 1F, 3F, 5F, 6F, 8F, and 9F. The fluxes recorded in the energy range 10--70~keV were summed over the seven detectors: 1F, 3F, 5F, 6F, 7F, 8F, and 9F. The 2F and 4F detectors were excluded due to their excessively enhanced backgrounds. All details concerning the processing of the \textit{RHESSI} data with the use of the OSPEX package of the SolarSoftWare (SSW) as the subtraction of the background from \textit{RHESSI} and \textit{GOES} data are described in detail in Paper I. For consistency with the results presented in Paper I, the whole X-ray flux was divided into four energy sub-ranges: 6--10, 10--20, 20--34 and 34--70~keV.

The reconstructed \textit{RHESSI} X-ray light curves have a native four-second-long time resolution determined by the rotation period of the spacecraft. To facilitate a qualitative comparison of concurrent variations of the emissions observed in the X-rays and the H$\alpha$ line, to obtain a much higher time resolution of 0.05 seconds, the X-ray light curves were demodulated to the time resolution of 0.25 seconds using the SSW demodulation procedure by \citet{2004GH}. All demodulated X-ray light curves were smoothed by a one-second-wide boxcar filter to suppress noises.

The X-ray spectra recorded by \textit{RHESSI} during the impulsive phase of the flare revealed thermal and nonthermal components. The spectra were therefore fitted with a single-temperature thermal plus thick-target (version 2) model (vth + thick2) available in the \textit{RHESSI} SSW software. The thick-target model was defined by the total integrated electron flux N\textsubscript{nth}, the power-law index of the electron energy distribution $\delta$, and the low-energy cut-off of the electron distribution E\textsubscript{c}. The actual values of the E\textsubscript{c} and $\delta$  were controlled and rectified in case of excessively abrupt variations. The fits were calculated in forward and backward directions, starting from a maximum of the impulsive phase, when the nonthermal component was strong and distinct. Figure~\ref{Fig06} shows three examples of the spectra collected during the pre-impulsive phase, the impulsive phase and just after the impulsive phase of the flare.


\begin{figure}[ht!]
\figurenum{7}
\begin{center}
\includegraphics[width=10.0cm]{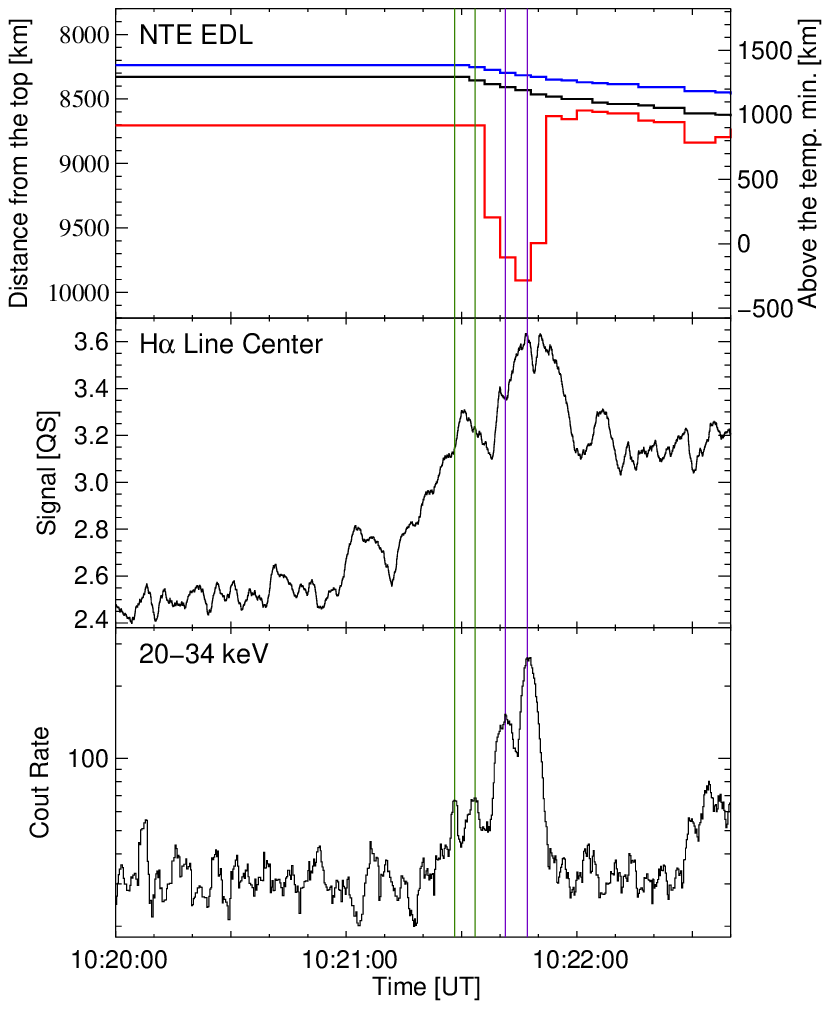}
\end{center}
\vspace{-0.3 cm}
\caption{Time variations of the numerically modeled position and the vertical extent of the energy deposition layer (EDL) and concurrently observed emissions in the hydrogen H$\alpha$ line center ($\lambda$~=~6563~{\AA}) and the hard X-ray emission in the 20--34~keV energy range. Vertical green and purple lines indicate four intermediate maxima of the hard X-ray emission. The H$\alpha$ light curve was smoothed with a one-second-wide boxcar filter and scaled to the level of the quiet Sun (QS). In the upper panel, the position of the peak of the energy flux deposited by the NTEs is shown with a black line, lower and upper limits of the EDL are shown with red and blue lines, respectively. The lower limit of energy flux in the EDL was selected arbitrarily as 0.01 erg s\textsuperscript{-1}cm\textsuperscript{-3}.
\label{Fig07}}
\end{figure}

The spectra-images collected with the \textit{MSDP} spectrograph were processed in a standard way, discussed in detail in previous publications by \citep{2006AdSpR..37.1317R, 2013SoPh..284..397R}. For all spectra-images, the H$\alpha$ line profiles were reconstructed for all pixels of the FOV in the spectral range of $\Delta$$\lambda$~=~$\pm$1.2~{\AA} from the H$\alpha$ line center. Furthermore, quasi-monochromatic images of the entire FOV were reconstructed in 13 wavelengths separated by 0.2~{\AA} (an effective waveband of the images was equal to 0.06~{\AA}). Due to some instabilities of the telescope pointing and a variable atmospheric seeing causing deformations and shifts of the observed structures, a special customized procedure was applied to correct displacements of H$\alpha$ images (see \citet{2007A&A...461..303R}, for details). The \textit{SDO}/HMI continuum images were used for precise coalignment of the H$\alpha$ images with the coordinate system of the satellite data. Using these data, the light curves of the flaring kernel were evaluated in freely selected wavelengths over the range $\Delta$$\lambda$~=~$\pm$1.2~{\AA} from the line center for the whole period of the \textit{MSDP} observations (Fig.~\ref{Fig07}).


\section{Impulsive phase of the flare} \label{sec:impuls}

In this section we present a detailed analysis of changes of H$\alpha$ hydrogen spectral line profiles ($\lambda$\textsubscript{0}=6562.8~{\AA}) during the impulse phase of the 2012 September 10 flare to relate fast changes of H$\alpha$ emission to energy conveyed to the chromosphere by NTEs. In general, variations of the integrated emission of the whole flare kernel, which can be delimited by an arbitrarily selected isophote, have significantly different time histories from the variations of the emission of selected, brightest nuclei of the kernel, limited also by another arbitrarily selected and substantially higher isophote. For purposes of this analysis, measurements of the integrated emissions of the flaring kernels and local nuclei of the kernel were performed applying isophotes of 90$\%$ and 98$\%$ of the highest signal, respectively. Here and everywhere further in this work, the emission level is referred to as a measured emission with subtracted emission of the quiet chromosphere.

Figures~\ref{Fig04}a and ~\ref{Fig04}b show variations of the mean profiles of the H$\alpha$ line emitted by the flaring kernel delimited by 90$\%$ isophote of the maximal emission, while Fig.~\ref{Fig05} shows variations of the integrated emission of the kernel delimited by the same isophote.

The \textit{MSDP} spectrograph provides spectral observations with a high time resolution (0.05 seconds), collected simultaneously for the entire observed region of the solar disk while maintaining 2-dimensional spatial resolution (limited by seeing only). Thanks to this, it was possible to determine a course of fast local variations of the H$\alpha$ line intensities and line profiles emitted by spatially limited, bright sub-nucleuses of the flaring kernel. Due to the high time correlation of variations of the emission observed in HXRs and the emission of the selected small nuclei of the flaring kernel in the H$\alpha$ line, these nuclei were considered as spatially limited and in time-limited regions of energy deposition by beams of the non-thermal electrons.
The conductive transport of the energy along the flaring loops has not been considered, because transport through conductivity is much slower, being typically 10 s \citep{2007A&A...461..303R, 2011A&A...535A.123R}, and so cannot be responsible for the fast changes seen.

\begin{figure}[ht!]
\figurenum{8}
\includegraphics[width=5.0cm]{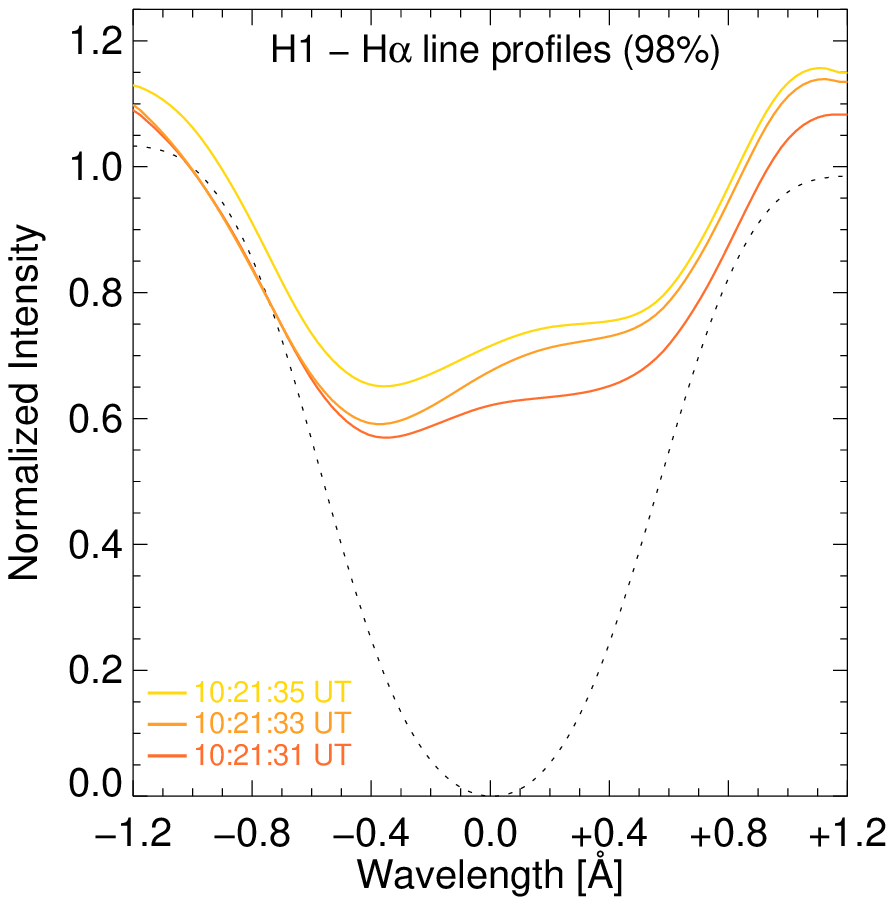}
\includegraphics[width=5.0cm]{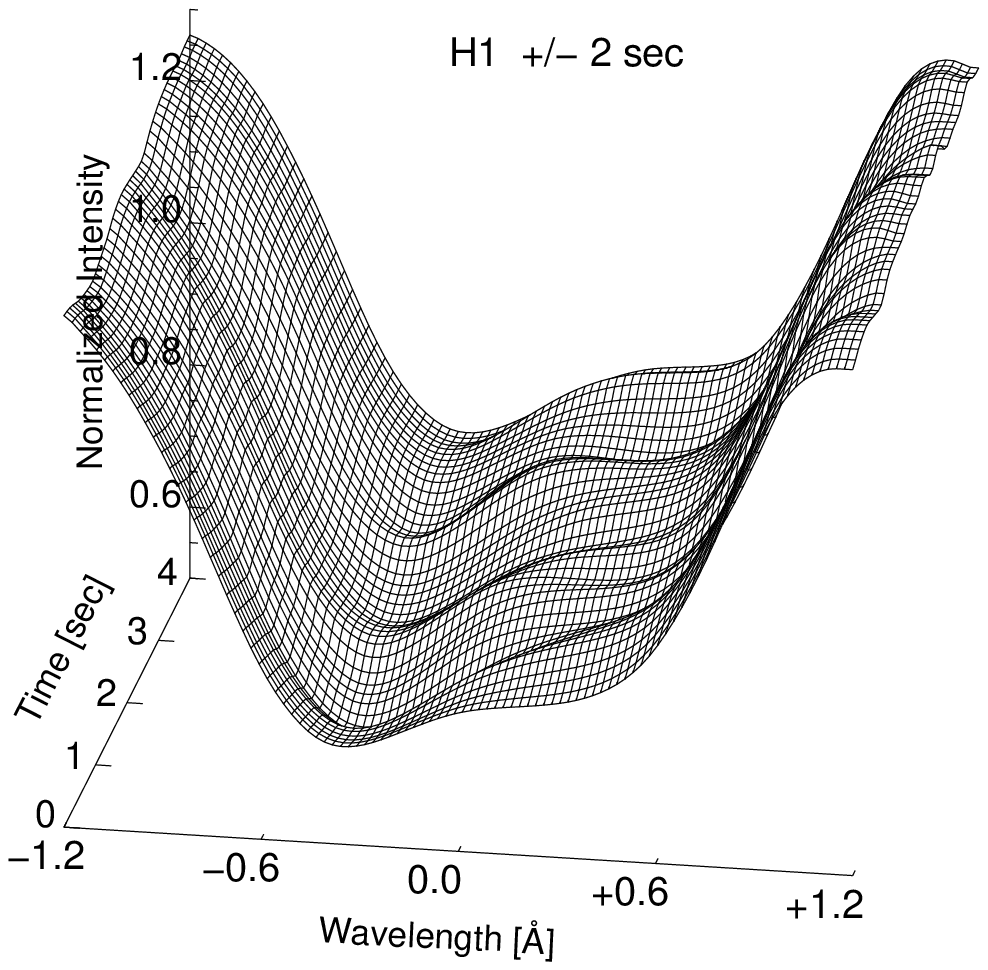}

\includegraphics[width=5.0cm]{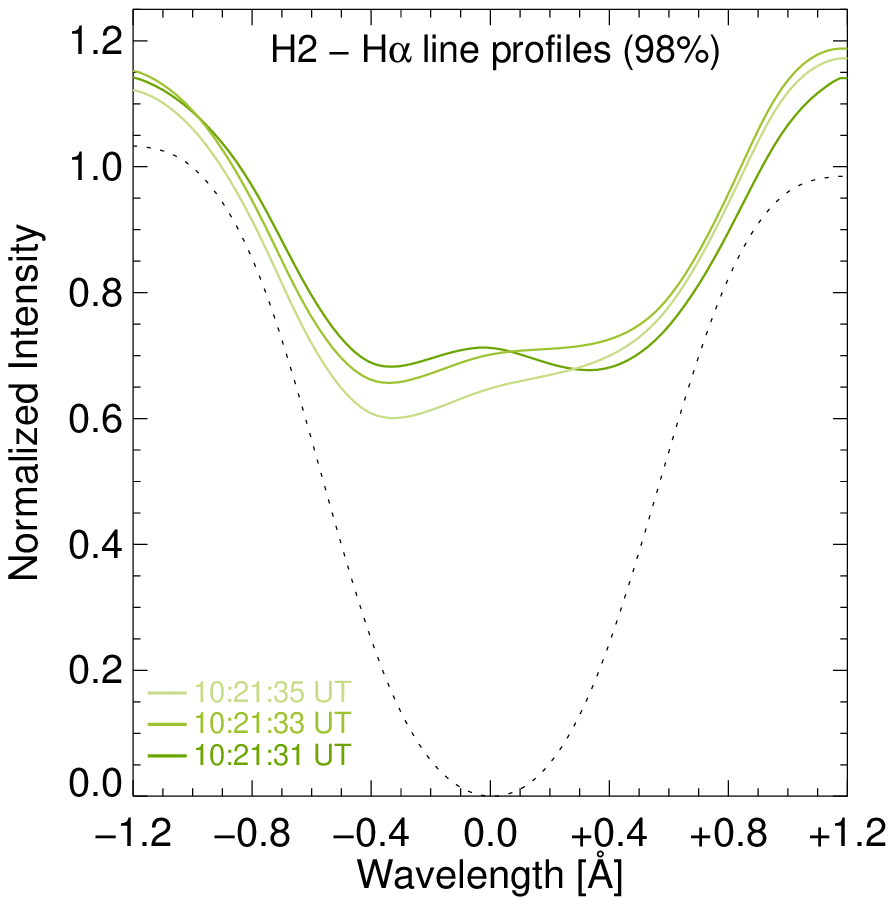}
\includegraphics[width=5.0cm]{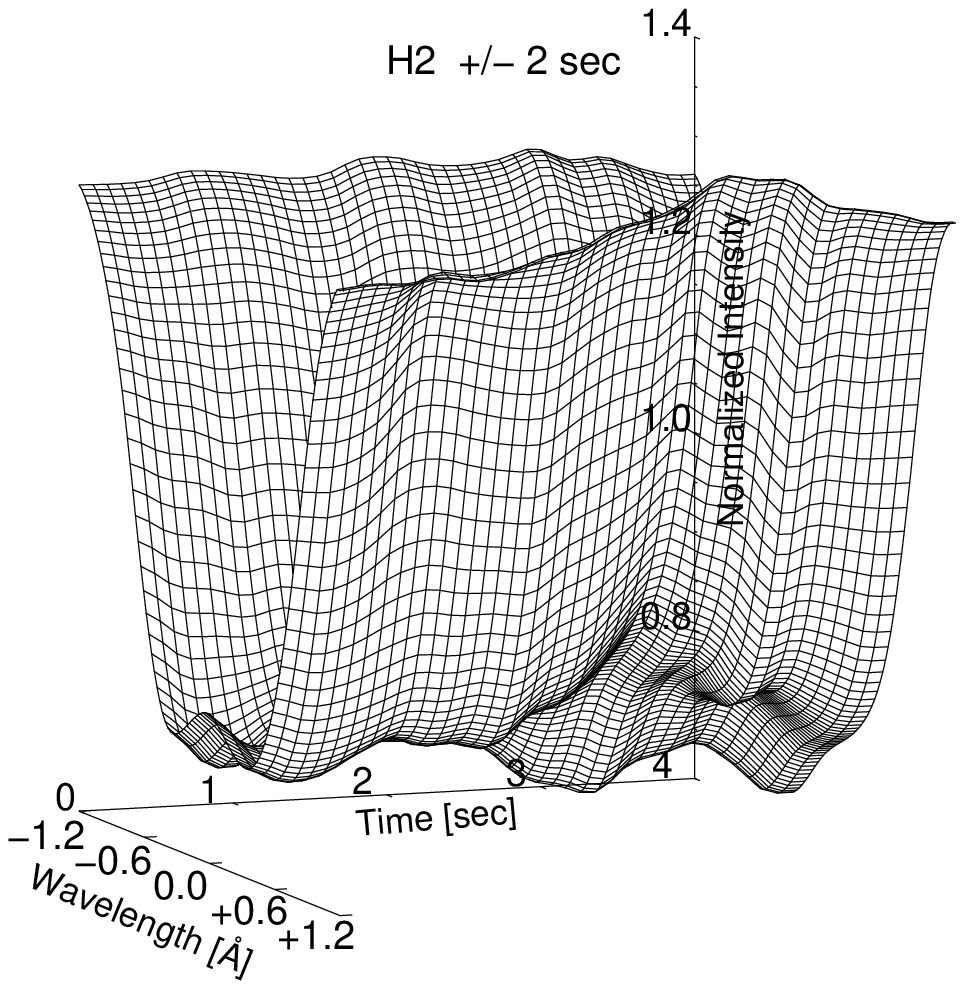}

\includegraphics[width=5.0cm]{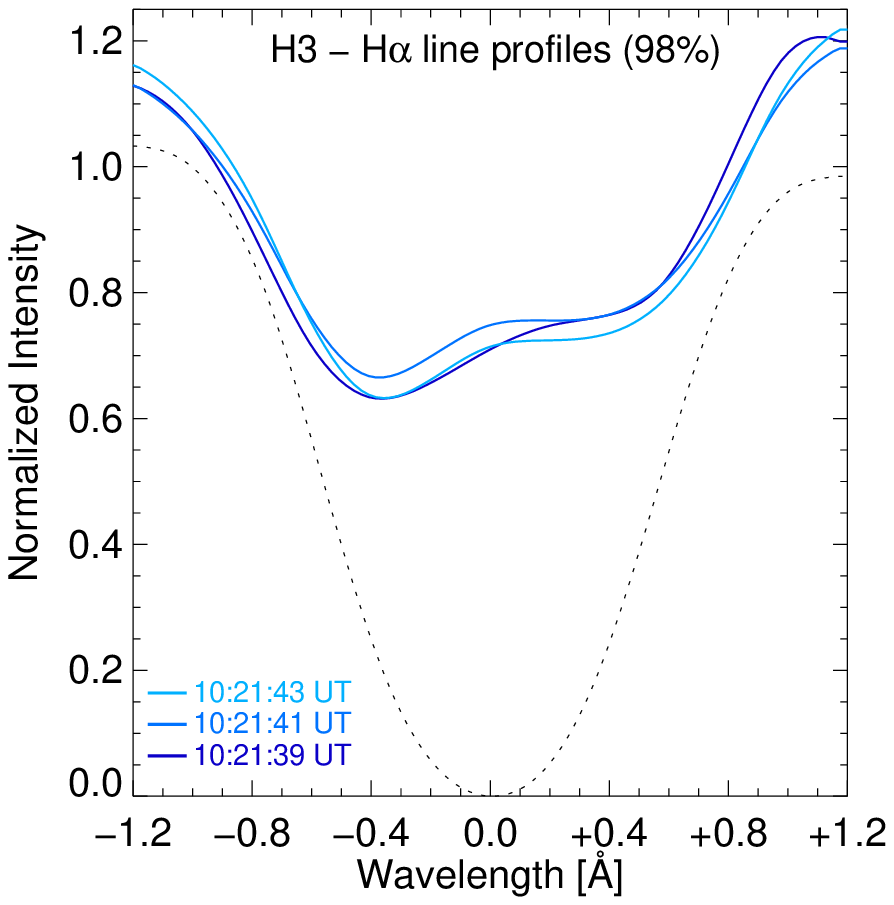}
\includegraphics[width=5.0cm]{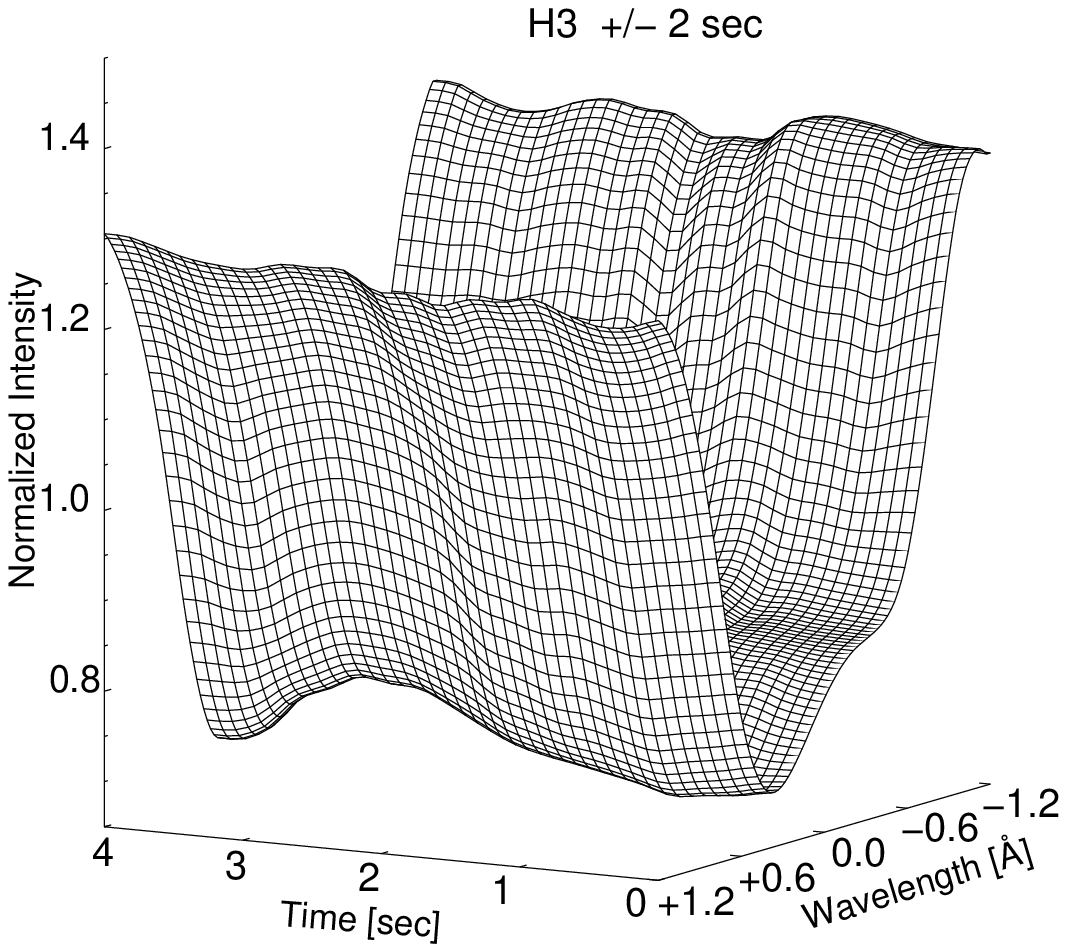}

\includegraphics[width=5.0cm]{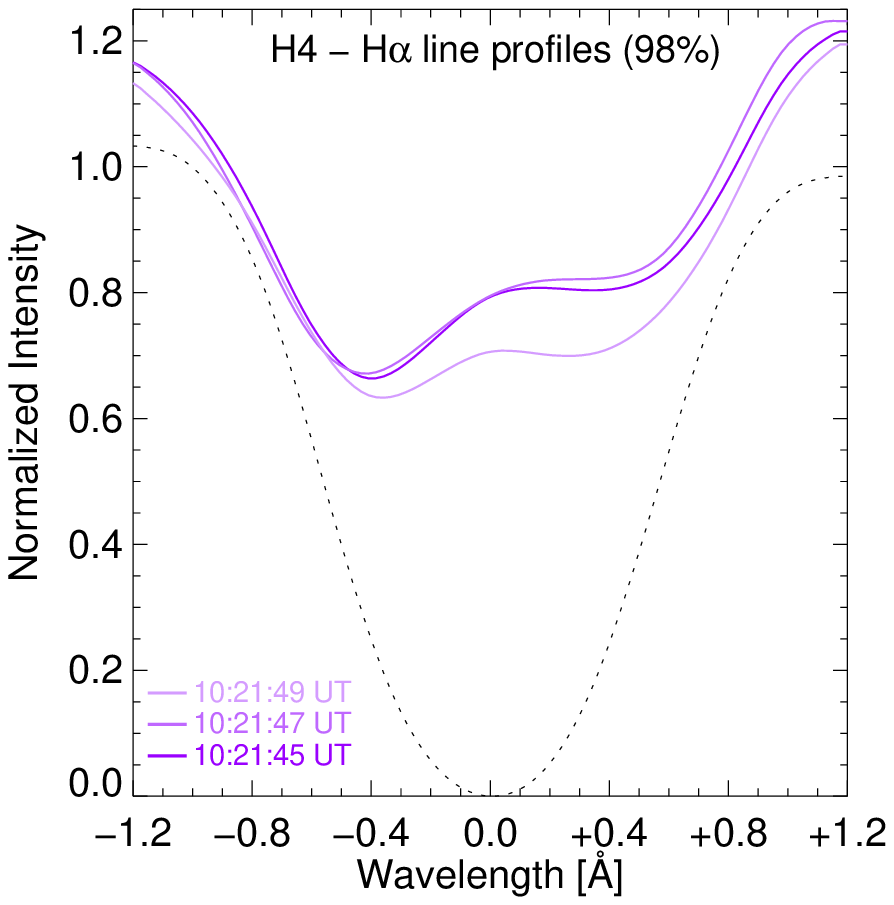}
\includegraphics[width=5.0cm]{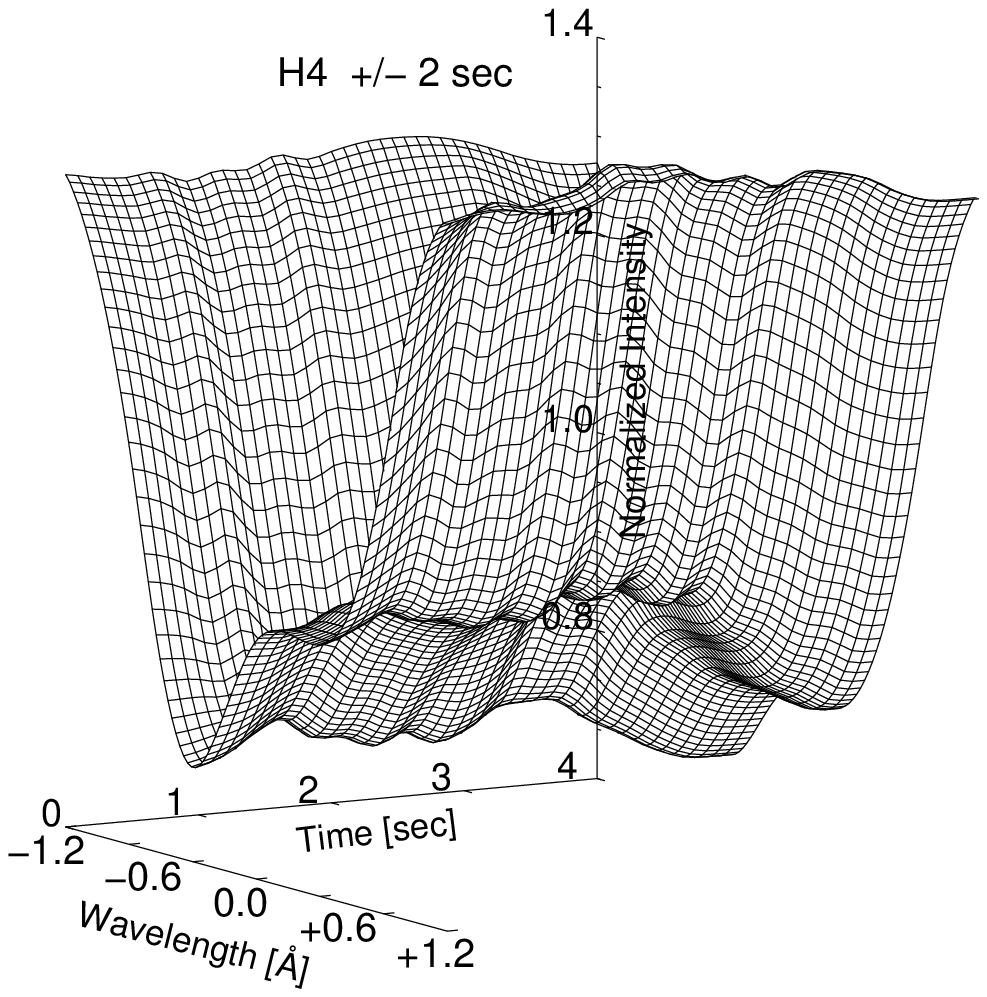}
\vspace{-0.3 cm}
\caption{Left column: H$\alpha$ line profiles of the brightest part of the flaring kernel delimited with an isocontour of 98$\%$ of the highest intensity in the H$\alpha$ line center recorded two seconds before, at the moment, and two seconds after the H1, H2, H3 and H4 peaks of the HXR recorded at 10:21:28 UT, 10:21:33 UT, 10:21:41 UT, and 10:21:47 UT, respectively. Right column: four-second-long series of the mean H$\alpha$ line profiles of the same part of the flaring kernel (98$\%$ -- like in the left column), and the same time intervals for H1--H4 peaks. The individual time-series of profiles in the right column are displayed at the right angle to highlight individual character of profile's changes. All profiles are normalized as in Fig.4.
An animation of the right column panels is available. In the video the H$\alpha$ line profiles are rotated about the Intensity axis to give views of the time and wavelength axes. The video makes on rotation in the 9 second duration.
\label{Fig08}}
\end{figure}

During the impulsive phase of the investigated flare four separate maxima (pulses) of the HXR emission were recorded at T\textsubscript{01}=10:21:28 UT, T\textsubscript{02}=10:21:33 UT, T\textsubscript{03}=10:21:41 UT, T\textsubscript{04}=10:21:47 UT and marked as H1, H2, H3, and H4, respectively (see Section 2, for details). For each of these pulses, the time evolution of the H$\alpha$ line profiles emitted from the lightest area in a frame of a whole H$\alpha$ flaring kernel was analyzed within a period DTi lasting from T\textsubscript{0i}-2 sec to T\textsubscript{0i}+2 sec, respectively. Due to a short duration of the non-thermal electron beams and a short reaction time of the chromosphere to the local deposition of the energy, the selected length of the time intervals is sufficient for this analysis (Fig.~\ref{Fig08}).

During the DT1 interval (H1 pulse), the mean H$\alpha$ line profile of the nucleus of the flaring kernel delimited by 98$\%$ isophote was continually asymmetrical, having emission in the red wing higher than in the blue wing. During the whole time-interval, a systematic but gradual increase of the H$\alpha$ line emission was observed, each step lasting about a second. The observed variations of the H$\alpha$ emission may indicate that the stream of the non-thermal electrons inducing the H1 pulse of HXR radiation was varied in time, showing sub-second-long changes in the efficiency of the magnetic reconnection and/or acceleration of electrons.

During the DT2 interval (which included the H2 pulse), very abrupt changes of the mean H$\alpha$ line profile of the nucleus of the flaring kernel delimited by 98$\%$ isophote were observed. In just half a second, from 10:21:32.75 UT to 10:21:33.25 UT, the emission in the blue wing of the profile at the wavelength $\lambda$ = $\lambda$\textsubscript{0}-0.4~{\AA} dropped sharply, but increased in the red wing of the line at the wavelength $\lambda$ = $\lambda$\textsubscript{0}+0.4~{\AA}. The strong asymmetry of the profile persisted until just over a second after T\textsubscript{02} (after the maximum of H2), and then it decreased, but the profile remained asymmetrical all the time. The strong asymmetry of the profile lasted less than 2 seconds and it was well correlated in time with the HXR pulse marked H2. Sudden changes of asymmetry of the profiles measured for the wavelengths $\lambda$ = $\lambda$\textsubscript{0}~$\pm$0.4~{\AA} can be attributed to short-lasting plasma motions with a velocity amplitude of  $\pm$20 km s\textsuperscript{-1}.

In a case of the H3 pulse, the emission of the sub-nucleus of the flaring kernel delimited by 98$\%$ isophote measured it the H$\alpha$ line, reached a maximum one second after the maximum of the H3 pulse in HXR (at 10:21:42 UT) and then began to decrease, so as little as two seconds after the impulse (at 10:21:43 UT), the emission measured in the center of the H$\alpha$ line returned to the pre-pulse level. In total, the impulsive enhancement of the H$\alpha$ emission lasted only three seconds, while the emission profile remained asymmetrical throughout.

In the case of the H4 pulse, during the DT4 period, the H$\alpha$ emission of the nucleus of the flaring kernel delimited by 98$\%$ isophote reached a maximum in less than one second after the maximum emission recorded in HXR (at 10:21:47.60 UT), and then began to decrease, reaching 1.5 s after the maximum of the HXR pulse (at 10:21:48.50 UT) a lower level than before the pulse. Such a fast decrease of the emissions in the H$\alpha$ line agrees with the assumption of a very fast relaxation of the chromosphere after a rapid heating by NTEs.

Figure~\ref{Fig08} shows variations of the mean H$\alpha$ profiles emitted by the relevant sub-nucleuses of the flaring kernel delimited by 98$\%$ isophote during the four-second time intervals encompassing H1, H2, H3, and H4 HXR pulses, respectively. Figure~\ref{Fig08} (left column) shows three representative H$\alpha$ profiles selected from the profiles shown in Figure~\ref{Fig08} (right column): the H$\alpha$ line profile at the moment of the HXR pulse maximum, as well as profiles measured 2 seconds before and 2 seconds after the maximum of HXR pulse. Four examples of H$\alpha$ line profiles evolution for various nuclei (98$\%$ isophote) of the flaring kernel are given in online material.

\section{Numerical model of the flare} \label{sec:model}

The 1D--HD numerical model of the flare was calculated using the modified hydrodynamic one-dimensional Solar Flux Tube Model (see \citet{1982ApJ...255..783M, 1989ApJ...341.1067M}, \citet{2009A&A...500..901F}, for details). A discussion of the use and accuracy of the 1D--HD numerical models of the solar flares was given in Paper I (\citet{2017ApJ...847...84F} and references therein), where we showed, that 1D--HD numerical models remain a valuable tool in investigations of the solar flares thanks to their moderate complexity and moderate volume of the necessary calculations.

The model was calculated under basic assumptions that plasma confined in the flaring loop was heated only by variable NTE beams. This assumption is supported by the results of \citet{2011ApJ...733...37F, 2014ApJ...789...71F} that in some flares the energy delivered by NTEs fulfills the energy budgets during the pre-impulsive and impulsive phases of the flare. However, in many other flares, the energy carried by NTEs is not large enough to balance the budget, thus some auxiliary energy sources and transfer mechanisms are active \citep{2013ApJ...770..111L, 2016ApJ...832...27A}.

The restored \textit{RHESSI} X-ray images were used for the evaluation of the main geometric characteristics of the flaring loop. Following Paper I, the area and positions of the flare footpoints were evaluated using the SSW CENTROID procedure, the 30$\%$  intensity contour was selected as the delimiter of the footpoints, and the perspective foreshortening was reduced using the method proposed by \citet{1999ApJ...517..977A}. To obtain a reasonable agreement between synthetic and observed X-ray light curves, the cross-section and the length of the loop were arbitrarily adjusted in their error ranges. The initial pressure in the transition region in the loop and the spatial distribution of temperature along the loop were selected to obtain a steady-state of the plasma inside the loop before the flare. The derived constant cross-section of the loop was equal to S~=~(1.14~$\pm$0.93)$~\times$~10\textsuperscript{16}~cm\textsuperscript{2},\ the half-length was equal to L\textsubscript{0}~=~(6.71~$\pm$1.13)~$\times$~10\textsuperscript{8}~cm under the assumption that the loop was semicircular. The following values of the cross-section and half-length were used in the model: S~=~2.27~$\times$~10\textsuperscript{16}~cm\textsuperscript{2}, L\textsubscript{0}~=~7.84~$\times~$10\textsuperscript{8}~cm, and the initial pressure in the transition region was equal to P\textsubscript{0}~=~35~dyn cm\textsuperscript{-2}. Following \citet{1994ApJ...434..786S} and \citet{2005psci.book.....A} a constant magnetic field of 200 G was assumed. A detailed discussion of the technical aspects of the model is given in Paper I.

The basic parameters of the NTE beams were derived from hard X-ray spectra recorded by the \textit{RHESSI}. The low cut-off energy E\textsubscript{c} was optimized so as to equalize the observed and synthetic \textit{GOES} 1--8~{\AA} fluxes by modification of an amount of the absorbed energy. Steady-state spatial and spectral distributions of the NTEs along the flaring loop were calculated for each time step of the model using the Fokker--Planck formalism \citep{1990ApJ...359..524M}. Using these data, spatial distributions of the thermodynamic parameters of the plasma -- X-ray thermal and nonthermal emissions, and the integral fluxes in the selected energy ranges -- were calculated for each time step. The thermal emission of the optically thin plasma was based on the X-ray continuum and line emissions calculated using the CHIANTI (version 7.1) atomic code \citep{1997A&AS..125..149D, 2006ApJS..162..261L}. For the plasma temperatures above 10\textsuperscript{5} K, the element abundances are based on the coronal abundances \citep{2000PhyS...61..222F}, while below 10\textsuperscript{5} K photospheric abundances were applied.

\begin{figure}[ht!]
\figurenum{9}
\begin{center}
\includegraphics[width=9.0cm]{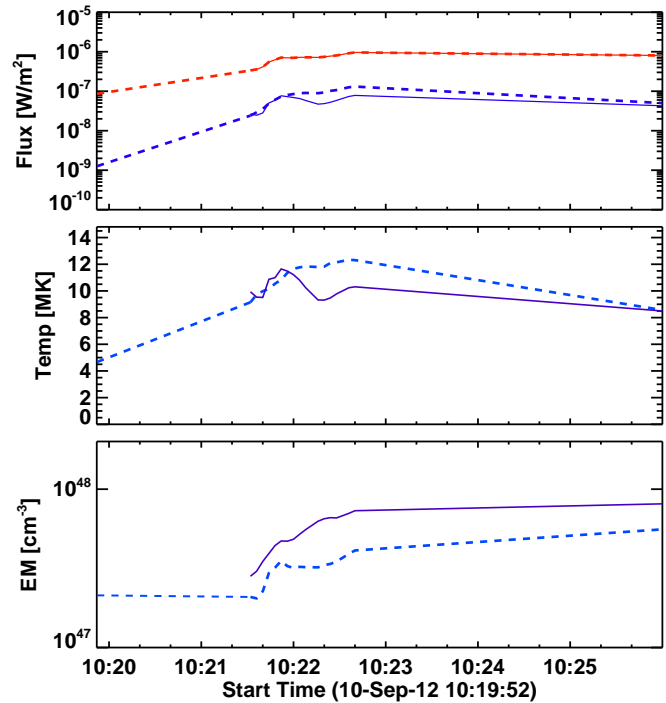}
\end{center}
\vspace{-0.3 cm}
\caption{Fluxes, temperatures and emission measures derived from \textit{GOES-15} data recorded before and during the impulsive phase of the 2012 September 10, flare compared with the same quantities calculated using the 1D--HD numerical model of the flare. Upper panel: X-ray fluxes recorded by \textit{GOES-15} in the 0.5--4~{\AA} and 1--8~{\AA} energy bands (dashed red and blue lines, respectively) and \textit{GOES} fluxes synthesized using the numerical model (solid red and blue lines). Middle panel: the temperatures derived from \textit{GOES} data (dashed light-blue line) and \textit{GOES }temperatures synthesized using the numerical model (solid dark-blue line). Lower panel: emission measures derived from \textit{GOES} data (dashed light-blue line) and \textit{GOES} emission measures synthesized using the numerical model (solid dark-blue line).
\label{Fig09}}
\end{figure}

Emitted X-ray fluxes, plasma temperatures and emission measures, derived from \textit{GOES-15} fluxes before and during the impulsive phase of the flare, compared with the same quantities calculated using the numerical model of the flare are shown in Figure~\ref{Fig09}. The computed \textit{GOES} temperatures are representative of the plasma with the highest differential emission measure in the flare. The emission measures were calculated as integrals of the computed differential emission measures for plasma hotter than 1 MK only. The synthetic \textit{GOES} 0.5--4~{\AA} flux calculated at various stages of the plasma in the numerical model followed qualitatively the observed one. The fluxes were nearly equal at the beginning of the flare only, between 10:21:30 UT to 10:22:00 UT, but after 10:22 UT up to 10:26 UT, the modeled flux was always lower than the observed one. The synthetic \textit{GOES} 1--8~{\AA} flux equaled the observed flux if the E\textsubscript{c} were adjusted. The slight differences between the observed and modeled \textit{GOES} 0.5--4~{\AA} fluxes (Fig.~\ref{Fig09}, upper panel) are reflected in time variations of the plasma temperatures and emission measures (Fig.~\ref{Fig09}, central and lower panels) because any variation of the 0.5--4~{\AA} to 1--8~{\AA} flux ratio causes proportional differences in the estimated temperatures and emission measures of the plasma \citep{2005SoPh..227..231W}. As a result, between 10:22 UT and 10:26 UT, the calculated temperatures were always lower, and the calculated emission measures were always higher than the respective observed quantities.


\section{Results} \label{sec:result}

The \textit{MSDP} spectrograph provides spectra with a high time resolution (0.05 seconds), simultaneously for the entire observed region of the solar disk while maintaining 2-dimensional spatial resolution (limited by seeing). It was thus possible to determine variations of the chromospheric flaring emission in the H$\alpha$ hydrogen line, resulting from short-lasting and spatially limited energy supplies by the beams of the non-thermal electrons. Local variations of the chromospheric emission in the H$\alpha$ hydrogen line during the impulsive phase of the compact solar flare of the September 10 flare were analyzed. The results of the observations collected in the H$\alpha$ line were compared with the measured HXR emission, while by demodulating the data the time resolution was improved to 250~ms \citep{2004GH}.

It has been found that the local changes of the H$\alpha$ emission, measured for the sub-nucleuses of the flaring kernel delimited by 98$\%$ isophote, are closely correlated in time with the variations of the HXR emissions. These variations occur significantly more dynamically than variations of the emission of the entire flaring kernels, limited by any significantly lower isophote, e.g. 90$\%$ of maximum emission. Thus, these areas can be considered as local and short-lasting areas of the energy deposition by the non-thermal electron beams.

The variations of the emission of the individual sub-nucleuses of the flaring kernel occur in sub-second time-scales, showing various forms of variability: a) pulsating increases in brightness (H1 pulse); b) sudden asymmetries of the H$\alpha$ profile if measured for $\lambda$ = $\lambda$\textsubscript{0}~$\pm$0.4~{\AA} (H2 pulse); c) impulsive increases of the emission (H3 pulse); d) abrupt but short-lasting increase of the emission while maintaining the asymmetry of the profile (H4 pulse).

Measurements of variations of a mean H$\alpha$ emission of entire flaring kernels are useful for analyzing the global features of the phenomenon (e.g. flash curve analysis) only. However, the analysis of individual episodes of local interactions between non-thermal electron beams and a chromosphere is possible only with the use of spectral observations made with high temporal and spatial resolution and should be carried out using data collected for selected, small areas only, with the highest emission within the whole H$\alpha$ flaring kernel.

The observed rapid local variations of the profiles and emission intensities of the H$\alpha$ line testify to very fast changes in the parameters of the chromospheric plasma as a result of rapid changes of the energy flux supplied locally by the beams of the non-thermal electrons during subsequent episodes of the impulsive phase. The course of variations of the emissions profiles of the H$\alpha$ line compared with the variations of the HXR emissions indicates that the reaction time of the chromosphere to changes in the supplied energy flux manifested through the changes in the profiles of the emitted H$\alpha$ line is only a few tenths of a second.

\begin{figure}[ht!]
\figurenum{10}
\begin{center}
\includegraphics[width=9.0cm]{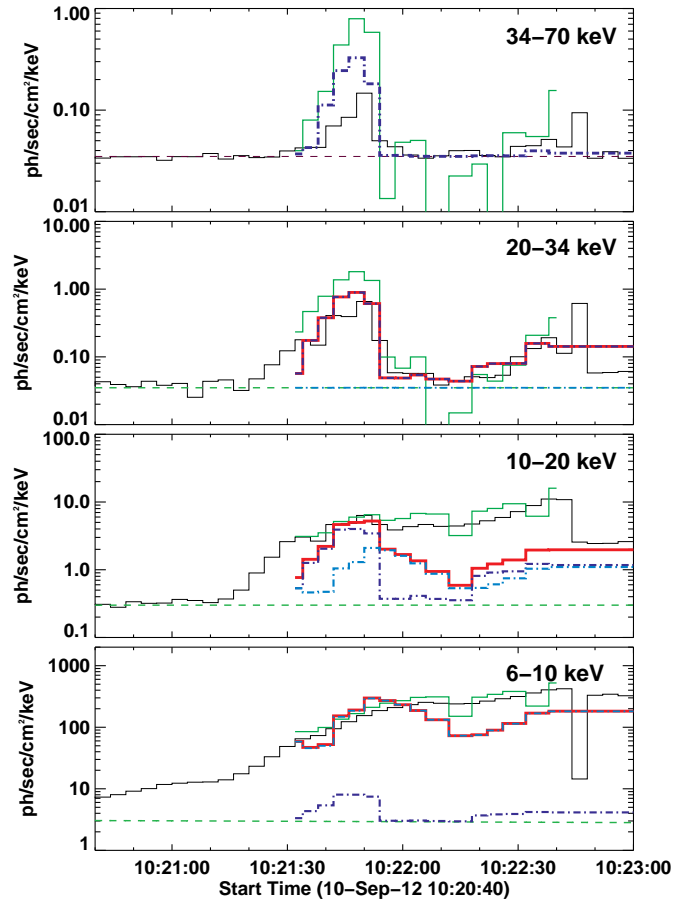}
\end{center}
\vspace{-0.3 cm}
\caption{X-ray fluxes recorded by \textit{RHESSI} satellite in the 6--10~keV, 10--20~keV, 20--34~keV, and 34--70~keV energy bands during the impulsive phase of September 10 flare, confronted with the synthesized X-ray fluxes calculated using the numerical 1D--HD model of the flare (see main text for details). The observed X-ray fluxes restored using the diagonal coefficients of the \textit{RHESSI}'s instrumental response matrix are shown by black lines, while the same fluxes restored using the full instrumental response matrix are plotted as green lines. The synthesized nonthermal and thermal X-ray fluxes are shown as dark blue and light-blue dotted lines, respectively. The sums of the synthesized thermal and nonthermal fluxes are shown as red lines. Dashed green lines indicate background levels.
\label{Fig10}}
\end{figure}

To compare the X-ray light curves observed by the \textit{RHESSI} and synthesized X-ray light curves calculated with the model, the calculated thermal and nonthermal emissions and the estimated observed background emission were summarized. For the most rapidly varying part of the impulsive phasee, between 10:21:30 UT and 10:23:00 UT, four light-curves summarized over 6--10~keV, 10--20~keV, 20--34~keV, and 34--70~keV energy bands are presented in Figure~\ref{Fig10}. While the calculated X-ray fluxes were expressed in photons cm\textsuperscript{-2} s\textsuperscript{-1} keV\textsuperscript{-1}, the observed X-ray fluxes were converted to the same units using two standard methods. In the first method, the \textit{RHESSI} instrumental response matrix with diagonal coefficients only was applied and the time resolution was equal to four seconds. In the second method, the full \textit{RHESSI} instrumental response matrix was used, but the time resolution was much worse, due to the much longer necessary accumulation time.

In the 34--70~keV and 20--34~keV energy bands, the X-ray emissions of the flare were dominated by non-thermal processes and thus the variations of the non-thermal component define the total emission. Due to obvious differences in the applied conversion methods, the converted observed fluxes in this energy range differ by a factor up to five, but the calculated variations of the X-ray flux are within the statistical ranges of the observed flux count rates. The relative contribution of the non-thermal component decreases in the lower energy bands, wherein the 6--10~keV band the thermal component plays a dominant role. The agreement of the observed and synthesized light curves in both energy bands (6--10~keV and 10--20~keV) was good (to the factor of two) only during the short period (10:21:30 UT and 10:21:50 UT) when the NTE beams precipitated along the loop and the plasma just started the sudden evaporation. In the 10--20~keV energy band, the total emission was dominated by non-thermal flux, while in the 6--10~keV band by thermal flux, but in both cases, the modeled fluxes have correct proportoins and total amounts. However, after this period the observed emission was dominated by thermal emission and the observed flux became one order larger than the calculated one, apparently due to the errors of the modeled densities and temperatures of the plasma along the loop.


The NTEs deposit the energy along the whole flaring loops, but a vast majority of the energy is deposited inside limited plasma volumes called the energy deposition layer, located chiefly in the chromospheric parts of the loops. The upper and lower boundaries of the EDL can be defined arbitrarily (following the Paper I) as altitudes where the deposited energy fluxes start to increase rapidly nearby the chromosphere and where they drop below 0.01 erg s\textsuperscript{-1} cm\textsuperscript{-3} inside the chromosphere, respectively.

The \textit{GOES} and the \textit{RHESSI} satellites recorded the first noticeable increase of the X-ray emission at 10:18:30 UT. The pre-flare gradual increase of the emission ended at about 10:21:12. An impulsive increase of the X-ray emission started at 10:21:12 UT in all applied \textit{RHESSI} energy channels. During the impulsive phase, the \textit{RHESSI} also recorded four individual intermediate peaks of the hard X-ray emission, well seen in energies above 10~keV. The first two pulses peaked in the energy range 20--34~keV at 10:21:28 UT and 10:21:33 UT, respectively. The two much stronger pulses peaked at 10:21:41 UT and 10:21:47 UT, respectively. Due to various share of the thermal emission, the hard X-ray emission above 20~KeV, returned to the pre-flare background level at about 10:21:53 UT, while in 6--10~keV and 10--20~keV bands the fluxes persistently increased beyond the end of the impulsive phase of the flare. During the early phase of the flare, the maximal flux of the deposited energy was of the order of 2839 erg s\textsuperscript{-1} cm\textsuperscript{-3}, causing gradual heating of the local plasma. The \textit{GOES}-class of the flare peaked at C1.6 only, but the flare emitted X-rays having energies up to 70~keV.

Up to 10:21:32 UT, the calculated vertical extent of the EDL equal to D~=~500~km remained constant due to the slow evaporation of the plasma (Fig.~\ref{Fig07}). The upper boundary of the EDL has initially an altitude of H\textsubscript{u}~=~1400~km above the temperature minimum. It started a slow, continual decline concurrently with the H2 intermediate peak of the X-ray emission at about 10:21:33 UT. As a result, up to 10:21:35 UT, the thickness of the EDL was gradually reduced to D~=~450~km and the altitude of its upper boundary dropped to H\textsubscript{u}~=~1350~km. Just after the end of H2 the thickness of the EDL abruptly increased to D~=~1150~km as a result of a sudden reduction of the lower boundary in response to a greatly increased flux of the deposited energy. At 10:21:40 UT and 10:21:44 UT, the EDL expanded again to D~=~1450~km and D~=~1625~km, respectively, responding to the energy deposited by the NTE beams manifested by the H3 and H4 peaks. At the moment of the maximum vertical extent, the upper boundary of the EDL was at an altitude of H\textsubscript{u}~=~1325~km while the lower boundary was at an altitude of H\textsubscript{D}~=~-300~km, i.e. below the temperature minimum level. Just after the maximum of the H4 intermediate peak, the lower boundary of the EDL returned instantly to the temperature minimum (H\textsubscript{D}~=~0~km) and just after the end of the H4 intermediate peak at 10:21:52 UT it returned instantly to the altitude of about H\textsubscript{D}~=~1000~km, a little above the altitude at the pre-impulsive phase of the flare. At the same time, the upper boundary of the EDL lowered gradually to H\textsubscript{u}~=~1250~km, thus the resulting vertical extent of the EDL shrank to D~=~250~km only. The minimal vertical extent of the EDL equal to D~=~175~km was achieved at 10:22:12 UT due to a persistent reduction of the altitude of the upper boundary and a small increase of the altitude of the lower boundary. At the same time also ended the impulsive enhancement of the H$\alpha$ emission. The EDL widened again to D~=~420~km starting from 10:22:28 UT, concurrently with the subsequent increase of the hard X-ray flux best seen in the 20--34~keV but noticeable also in remaining X-ray energy bands.

Due to the perspective effects, the H$\alpha$ brightening was located near the central part of the extended loop-shaped structure visible in the X-rays images (see Fig.~\ref{Fig01}). The increased H$\alpha$ emission was recorded from the very beginning of the observations at 10:20:00 UT in the red wing of the H$\alpha$ line (see Fig.~\ref{Fig05}) and about 10:20:50 UT in the H$\alpha$ line center. In the blue wing of the line profile, the emission dropped a little up to 10:21:10 UT but sharply increased afterwards. The light curves presented in Figure~\ref{Fig05} are scaled to the signal of the quiet Sun (QS), and the wavelengths of the emission are measured in the reference systems of the emitting plasma, i.e. correcting  for solar rotation and proper motions along the flaring loop.

An intermediate peak of the H$\alpha$ emission, recorded between 10:21:30 and 10:21:33 UT in the blue wing of the line, at 10:21:30 UT in the line center and 10:21:33 UT in the red wing occurred concurrently with the H1 and H2 peaks of the X-ray emission recorded by the \textit{RHESSI} satellite. The H1 and H2 peaks had similar fluxes in 10--20~keV and 20--34~keV energy bands, but in the 34--70~keV energy band, the H1 peak was much fainter than the H2 peak. Thus, the NTE beams related to H1 and H2 peaks had various energies, and the NTEs related to the H1 peak precipitated a little shallower than the NTEs related to the H2 peak and they deposited their energy in various altitude ranges of the chromosphere. The maximum of the H$\alpha$ emission was recorded between 10:21:40 UT and 10:21:53 UT. The emission in the line center and the blue wing of the profile (-1.0~{\AA}) peaked almost simultaneously at 10:21:45 UT, while the maximum in the red wind (+1.0~{\AA}) peaked about 10:21:53 UT. The maximum occurred concurrently with the H3 and H4 peaks of the X-ray emission recorded by the \textit{RHESSI} satellite. The H3 peak of the HXRs had a substantially lower flux than the H4 peak in all energy bands, but their energy spectra were similar. Thus, the NTE beams deposited energy in both cases in the same range of altitudes and caused nearly simultaneous increases the emission in the whole profile.

\begin{figure}[ht!]
\figurenum{11}
\begin{center}
\includegraphics[width=9.0cm]{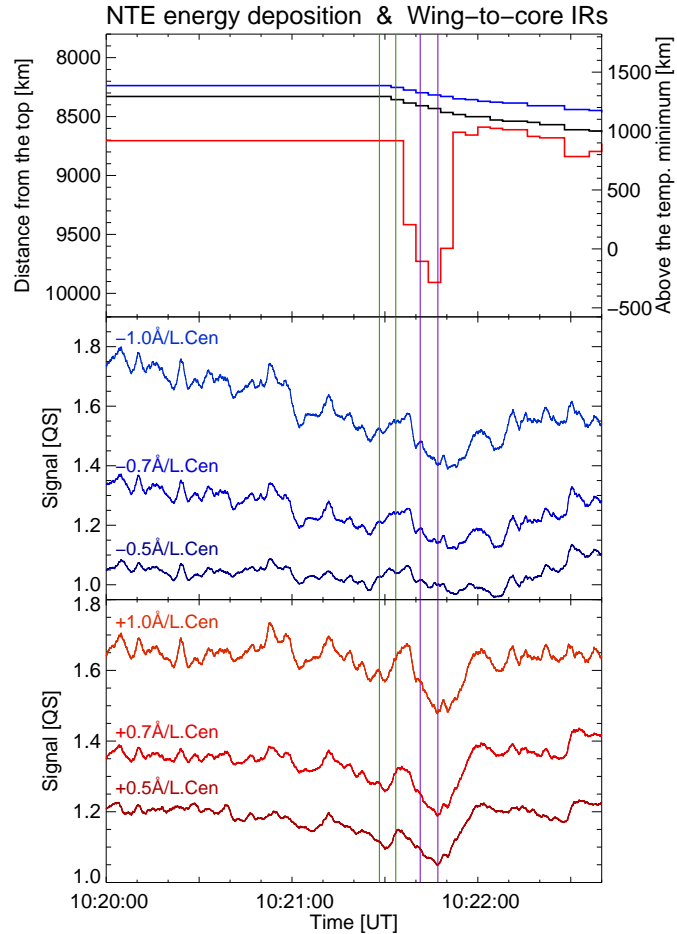}
\end{center}
\vspace{-0.3 cm}
\caption{Time variations of the numerically modeled position and extent of the EDL (upper panel) and time variations of the wing-to-line-center intensity ratios (IRs) of the H$\alpha$ flaring kernel in blue and red wings of the line at $\Delta$$\lambda$~=~$\pm$0.5~{\AA}, $\Delta$$\lambda$~=~$\pm$0.7~{\AA}, and $\Delta$$\lambda$~=~$\pm$1.0~{\AA} (middle and lower panels, respectively) during the impulsive phase of the C1.6 \textit{GOES}-class solar flare on 2012 September 10. Vertical green and purple lines indicate four intermediate maxima of the hard X-ray emission recorded by \textit{RHESSI} in the 20--34~keV energy range.
\label{Fig11}}
\end{figure}

The variations of the H$\alpha$ emission in various parts of the line profile are whown in Figure~\ref{Fig11} by the variations of the wing-to-line center intensity ratios (IRs). The IRs measured at $\Delta$$\lambda$~=~$\pm$0.5~{\AA}, $\Delta$$\lambda$~=~$\pm$0.7~{\AA}, and $\Delta$$\lambda$~=~$\pm$1.0~{\AA} measured for the period between 10:20:00 UT and 10:22:40 UT are shown in Figure~\ref{Fig11}, where the wavelengths are corrected for Doppler effect, i.e. the influence of the own plasma motions is removed. During the pre-impulsive phase of the flare, a slow, gradual decrease of the IRs occurred in all discussed wavelength ranges, most noticeable in the far wings of the line ($\pm$1.0~{\AA}). Concurrently with the first intermediate X-ray peak H1, the IRs peaked briefly at 10:21:28 UT. Following this, the IRs increased much more between 10:21:30 UT and 10:21:46 UT. During this increase, the relatively faint H2 intermediate X-ray peak occurred in a course of an ascent stage while a much stronger H3 peak occurred during a descending phase. Despite relatively high emission of the H3 peak, it was barely marked with a tiny intermediate increase of the IRs only.

The H4 intermediate peak of the X-ray emission, having the highest emission X-ray flux, occurred just at the beginning of a new strong increase of the IRs at 10:21:48 UT. The increase of the IRs was best seen in the red wing of the H$\alpha$ line, but it was also recognizable in the blue one. The IRs increased first to the level similar to the previous level, then after a decrease it remained constant or slightly increased at all wavelengths. The IR variations were synchronized with the relevant peaks of the X-rays and the time delays between X-ray impulses and abrupt variations of the H$\alpha$ profiles were virtually equal to zero.

\section{Discussion and Conclusions} \label{sec:disc}

The 1D hydrodynamic numerical model of the 2012 September 10 flare was calculated under the assumption that the NTEs carried the entire flare energy. The flare emitted hard X-ray flux detected by the \textit{RHESSI} up to an energy of 70~keV, thus the clear non-thermal component of the \textit{RHESSI} spectra was easily distinguished and the NTE energy derived. The model was optimized by an automatic fit of an actual value of the cut-off energy parameter E\textsubscript{c} to obtain a good match between observed and calculated \textit{GOES-15} soft X-ray fluxes measured in the 1--8~{\AA} waveband. The chief geometrical parameters of the flare, like loop length and diameter, were derived using derived imaging \textit{RHESSI} data. The modeled variations of the vertical extent and the altitude of the energy deposition layer were compared with variations of the high-time resolution spectra and emission intensities observed in the H$\alpha$ line.


To compare the X-ray light curves observed by the \textit{RHESSI} and synthesized X-ray light curves calculated with the numerical model of the flare, the emitted X-ray fluxes and the estimated backgrounds were calculated for the impulsive phase of the flare in 6--10~keV, 10--20~keV, 20--34~keV, and 34--70~keV energy bands (selected consistently with the energy bands applied in Paper I). The relevant observed X-ray fluxes were converted to photon cm\textsuperscript{-2} s\textsuperscript{-1} keV\textsuperscript{-1} units twice, firstly using the \textit{RHESSI} instrumental response matrix with diagonal coefficients for data with four-second time resolution, and secondly using the full \textit{RHESSI} instrumental response matrix for data with longer accumulation times.

The X-ray emission of the flare above 20~keV was dominated by the non-thermal processes and the calculated X-ray fluxes in the 20--34~keV and 34--70keV energy bands are within the limits of the observed fluxes restored by both methods (Fig.~\ref{Fig09}). Below 20~keV the relative contribution of the non-thermal component decreases substantially, and in the 6--10~keV band, the thermal component plays a dominant role. The agreement of the observed and synthesized light curves in 6--10~keV and 10--20~keV energy bands was good only during the short period when the NTE beams precipitated along the loop and the plasma just started the violent evaporation. During this period, in the 10--20~keV energy band, the emission was dominated by the non-thermal flux, while in the 6--10~keV band by the thermal emission. In both cases, the modeled fluxes have correct contributions and total amounts. Four intermediate peaks H1-H4 of the hard X-ray emission occurred at nearly even intervals in time. Such quasi-periodicity of the peaks could reflect a quasi-periodic variation of efficiency of an involved acceleration process of the non-thermal electron beams \citep{1994ApJS...90..631A, 2002ApJ...574L.101W, 2008MNRAS.388.1899S, 2010SoPh..263..163Z, 2019ApJ...875...33H}.

After the impulsive phase, the total observed emission of the flare was dominated by a thermal component. The observed flux was one order larger than the calculated one. As was discussed in Paper I, the above-mentioned differences between the modeled and the observed fluxes were caused by the inevitable discrepancies between the spatial distribution of the local thermodynamic and kinematic parameters of the plasma in the numerical model and inside the real flaring loop as well as by differences in precipitation depths of the NTEs of various energies. The NTE beams containing large populations of high-energy electrons (i.e., with a hard spectrum) effectively precipitate down to relatively low altitudes into the chromosphere because they weakly interact with the plasma confined in the upper part of the loop. As a result, the relatively small portion of the transmitted energy is deposited in the upper part of the chromosphere and/or transition region, causing moderate "gentle evaporation". In contrast to them, the low energy NTEs deposit efficiently their energy in the upper part of the chromosphere, where the densities are relatively low, giving rise to "explosive evaporation" (see \citet{2009A&A...500..901F} for details). Thus, the HXR non-thermal emission is directly related to the total flux of the accelerated NTEs, whereas the SXR and HXR thermal emissions are related to the energy effectively deposited by the NTEs. As was discussed in Paper I the assumed power-law energy spectrum of the injected NTEs also caused some differences between the modeled and the observed X-ray fluxes in the low-energy X-ray bands.

The variations of altitude and the vertical extent of the EDL are compared with the variations of the spectra and emission intensities recorded in the H$\alpha$ line using the Multi-channel Subtractive Double Pass imaging spectrograph in the Bia{\l}k{\'o}w Observatory. The spectra-images were recorded with the time cadence of 0.05 sec in an effective waveband of 2.4~{\AA}. High-cadence series of the quasi-monochromatic images of the whole FOV and the light curves of the selected flaring kernels at various wavelengths were derived.

In the September 10 flare, the H$\alpha$ emission of the flare was well correlated spatially and in time with the impulsive brightenings recorded in the hard X-ray by the \textit{RHESSI.} The H$\alpha$ emission reached a maximum during the impulsive phase of the flare, while the noticeable intermediate maxima of the H$\alpha$ emission coincided with intermediate emission peaks detected in the hard X-rays during the impulsive stage. Agreement of the spatial and time variations of X-ray fluxes at all energies with the H-alpha variations validate our previous results \citep{2007A&A...461..303R, 2011A&A...535A.123R, 2017ApJ...847...84F}.

The NTEs deposited the energy along the whole flaring loop, but a vast majority of the energy was deposited inside limited plasma volumes called the energy deposition layers, located in the chromospheric sections of the loops. The upper and lower boundaries of the EDL can be defined arbitrarily (following Paper I) as altitudes where the deposited energy fluxes start to increase rapidly nearby the chromosphere (the upper limit) and where they drop below 0.01 erg s\textsuperscript{-1} cm\textsuperscript{-3} inside the chromosphere (the lower limit), respectively. The vertical extent and the altitude of the EDL can be modeled numerically but cannot be observed directly in any spectral range. However, the spatial and time variations of the modeled emissions can be compared with the observed ones, indirectly confirming the correctness of the model and thus the correctness of the reconstructed parameters of the EDL and their changes.

The emission of the flare in the H$\alpha$ line started to brighten exactly at the beginning of the impulsive phase, at 10:20:00 UT in the red wing of the H$\alpha$ line (see Fig.~\ref{Fig04}), at about 10:20:50 UT in the H$\alpha$ line center, and at 10:21:10 UT in the blue wing of the line profile (after a short but noticeable drop of the intensity, see Section 6).

The vertical extent of the EDL during the initial stage of the impulsive phase was 500~km and it remained constant up to 10:21:32 UT due to the very slow evaporation of a gently heated plasma. An intermediate peak of the H$\alpha$ emission was recorded between 10:21:30 UT and 10:21:33 UT, concurrently with the H1 and H2 peaks of the X-ray emission recorded by the \textit{RHESSI} satellite. The NTEs beams related to the H1 and H2 peaks had various energies and they deposited its energy at various altitudes in the chromosphere.

At 10:21:44 UT the EDL expanded to D~=~1625~km, due to a sudden decrease in height of its lower boundary caused by substantially increased energy deposition at low altitudes in the chromosphere by the high-energy and deeply penetrating NTEs. These NTEs were manifested by the intermediate H3 and H4 HXR peaks well seen in the 34--70~keV energy bands. Just after the H4 peak, at about 10:21:53 UT, the vertical extent of the EDL shrank to D~=~250~km, because the energy flux deposited at low altitudes nearly vanished. The maximum of the H$\alpha$ emission was recorded at the same time, between 10:21:40 UT and 10:21:53 UT. The emissions in the H$\alpha$ line center and the blue wing of the profile (-1.0~{\AA}) peaked almost simultaneously at 10:21:45 UT, while the emission in the red wind (+1.0~{\AA}) peaked at about 10:21:53 UT. The H3 and H4 peaks of the HXRs had similar energy spectra, so the related NTEs electron beams deposited energy inside the same range of altitudes, down to an altitude of H\textsubscript{D}~=~200~km below the pre-flare altitude of the temperature minimum, and caused nearly simultaneous increases of the whole line profile.

The upper boundary of the EDL had initially the altitude of H\textsubscript{u}~=~1400~km above the temperature minimum and due to the plasma evaporation, it gradually reduced to H\textsubscript{u}~=~1250~km at the end of the HXR emission at 10:21:53 UT. Eventually it reduced to an altitude of H\textsubscript{u}~=~1175~km at the end of the modeled period as a result of an ongoing gentle plasma heating and evaporation by low-energy NTEs. The lower boundary of the EDL had initially an altitude of H\textsubscript{u}~=~900~km above the temperature minimum but due to three consecutive episodes of the intense heating by the high-energy NTEs, it suddenly decreased to an altitude of H\textsubscript{u}~=~-300~km at the maximum of the HXR emission at 10:21:44 UT, below the initial altitude of the temperature minimum. Just after the end of the HXR emission, the lower boundary of the EDL returned instantly to the altitude of about H\textsubscript{D}~=~1000~km, a little above the altitude noticed during the pre-impulsive phase of the flare.

The differences between variations of the emissions recorded in various parts of the H$\alpha$ line profile are well illustrated by the variations of the wing-to-line center intensity ratios (IRs). A slow, gradual decrease of the IRs occurred in all discussed wavelength ranges during the pre-impulsive phase of the flare. Next, the IRs increased noticeably between 10:21:30 UT and 10:21:46 UT, concurrently with the H2 intermediate HXR peak.  However, the much stronger H3 intermediate X-ray peak was barely marked with a tiny intermediate increase of the IRs only. The H4 intermediate HXR peak, having the highest X-ray flux, occurred simultaneously with a new and strong increase of the IRs at 10:21:48 UT. The increases of the IRs were best seen in the red wing of the H$\alpha$ line, but they were also recognizable in the blue one. The variations of the IRs were synchronized with the relevant peaks of the X-rays and the time delays between X-ray impulses and abrupt variations of the H$\alpha$ profiles were virtually equal to zero.

As a result, we found that the variations of the X-ray fluxes recorded in various energy bands and the variations of the H$\alpha$ intensities measured in various parts of the line profile were well correlated in time during the pre-impulsive and impulsive phases of the flare and they agreed well with the variations of the calculated position and variations of the vertical extent of the EDL. The variations of the emission noticed in various parts of the H$\alpha$ line profile were caused by individual episodes of the energy deposition by the NTEs of various energy spectra on various depths in the chromospheric plasma. Namely, the NTE beams manifested by hard X-ray emission up to 70~keV deposited their energy at low altitudes in the chromosphere, where the wings of the H$\alpha$ line are formed and varied concurrently with the increases of the wing-to-line center intensity ratio (IRs). The NTE beams manifested by relatively soft X-ray emission deposited their energy at much higher altitudes in the chromosphere, concurrently with the increases of the emission in the center of the H$\alpha$ line. These conclusions can be compared with our results for the solar flare on June 21, 2013, presented in detail in Paper I. The flare had nearly the same \textit{GOES}-class of C1.1 as the flare on 2012 September 10, but the \textit{RHESSI} detected HXR emission only below 34~keV. Both flares were investigated by us using the same methodology and similar observational data, and they were modeled with the same numerical code. The variations of the IRs ratios in a course of the impulsive phase of a solar flare indicate not only a relative variation of the intensities recorded in various parts of the line profile, but also variations of an actual effective altitude of the energy deposition layer.

In the case of the 2013 June 21 flare, the time variations of the H$\alpha$ line profiles emitted by flaring kernels were also caused by temporal variations of penetration depths of NTE beams along the flaring loops. However, due to much lower energies and penetration depths of the NTEs manifested by HXRs below 34~keV, the energy delivered to low altitudes in the chromosphere was low and the deposition period was short. As a result, IRs decreased abruptly just after the first impulse seen in the 20--34~keV energy band (see Fig. 13 in Paper I). Some short-lasting secondary increases of the IRs were correlated in time with the subsequent X-ray impulses recorded in both HXR and SXR ranges, related to secondary minor episodes of the energy deposition. Most of the energy carried by low-energy NTEs was deposited at relatively high altitudes of the chromosphere. In the case of the flare on 2012 September 10, a substantial amount of the energy was deposited by the high-energy NTEs at low altitudes.

In the case of both flares, the very fast variations of plasma properties in the lower parts of flaring loops are well revealed by the numerical models. The time scales of variations of the plasma parameters are compatible with the time scales of the variations of the chromospheric emission. The cross-comparison of the variations of the emissions recorded in various ranges of the spectrum (like X-rays and visual domain) validate the results of the numerical models and it allows estimations of these properties of the flaring plasma, which are not directly observable, for example, the altitude, the vertical extent and the time variations of the energy deposition layer in the feet of the flaring loop.


\section{Acknowledgments} \label{sec:ack}

 The authors acknowledge the \textit{RHESSI} and \textit{SDO} consortia for providing valuable observational data. The numerical simulations were carried out using the resources provided by the Wroc{\l}aw Centre for Networking and Supercomputing (http://wcss.pl), grant No. 330.

\bibliography{bibliography}{}
\bibliographystyle{aasjournal}

\end{document}